\documentclass[reprint,amsmath,amssymb,aps,floatfix,showpacs]{revtex4-1}
\usepackage{graphicx}
\usepackage{dcolumn}
\usepackage{bm}

\begin{document}

%\preprint{APS/123-QED}

\title{Intracavity Rydberg atom electromagnetically induced transparency using a high finesse optical cavity}

\author{Jiteng Sheng}
\author{Yuanxi Chao}
\author{Santosh Kumar}
\author{Haoquan Fan}
\author{Jonathon Sedlacek}
\author{James P. Shaffer}

\affiliation{Homer L. Dodge Department of Physics and Astronomy,The University of Oklahoma, \\440 W. Brooks Street, Norman, OK 73019, USA}

\date{\today}

\begin{abstract}
We present an experimental study of cavity assisted Rydberg atom electromagnetically induced transparency (EIT) using a high-finesse optical cavity ($F \sim 28000$). Rydberg atoms are excited via a two-photon transition in a ladder-type EIT configuration. A three-peak structure of the cavity transmission spectrum is observed when Rydberg EIT is generated inside the cavity. The two symmetrically  spaced side peaks are caused by bright-state polaritons, while the central peak corresponds to a dark-state polariton. Anti-crossing phenomenon and the effects of mirror adsorbate electric fields are studied under different experimental conditions. We determine a lower bound on the coherence time for the system of $7.26 \pm 0.06 \,\mu$s, most likely limited by laser dephasing. The cavity-Rydberg EIT system can be useful for single photon generation using the Rydberg blockade effect, studying many-body physics, and generating novel quantum states amongst many other applications.
\end{abstract}

\pacs{36.90+f, 39.25+k, 32.10-f, 33.80.Rv}

\maketitle

\section{INTRODUCTION}

Rydberg atoms, i.e. highly excited atoms corresponding to large principal quantum number, $n$, have been well studied for several decades \cite{gallagher2005rydberg}. At least in the case of alkali atoms, Rydberg atoms are now emerging as a tool for quantum technologies partly because their properties can be engineered by state selection and the application of electro-magnetic fields. Rydberg atoms possess many exaggerated properties that can be useful for controlling matter and electro-magnetic fields at the quantum level, such as large geometrical size, long lifetime, large transition dipole moments between neighboring levels, and large polarizability. Recently, research has focussed on their strong interactions, leading to the Rydberg blockade effect that allows collective excitations to be created \cite{tong2004local,heidemann2007evidence}. A number of review articles exist on various specific topics, e.g., Rydberg atom interactions \cite{comparat2010dipole,marcassa2014interactions}, quantum information with Rydberg atoms \cite{saffman2010quantum,saffman2016}, Rydberg atoms in magnetic fields \cite{pohl2009cold} and microwave field sensing with Rydberg atoms \cite{fan2015atom}. New experiments and theory, where collective Rydberg excitations created in ultracold gases are used to shape electro-magnetic fields at the quantum level are beginning to attract increasing attention \cite{gorshkov2011photon,dudin2012strongly,peyronel2012quantum,pritchard2012correlated,firstenberg2013attractive,he2014two,tiarks2014single,maghrebi2015coulomb,grankin2016inelastic,gorniaczyk2016enhancement,jachymski2016three,Grangier16,Thompson17}.

Cavity quantum electrodynamics (CQED) is aimed at investigating the interaction between matter and electromagnetic fields confined within a resonator and has been widely studied in a large variety of systems. The matter can be neutral atoms \cite{Kimble99,Berman1994}, ions \cite{sterk2012photon,casabone2015enhanced}, molecules \cite{tischler2005strong}, quantum dots \cite{reithmaier2004strong}, nitrogen-vacancy centers \cite{park2006cavity}, etc., and the resonator includes millimeter-wave cavities \cite{haroche1985radiative}, optical cavities \cite{thompson1992observation}, microtoroid cavities \cite{aoki2006observation}, photonic crystal defect cavities \cite{yoshie2004vacuum}, fiber cavities \cite{haas2014entangled}, superconducting stripline resonators \cite{blais2004cavity}, surfaces \cite{Sheng2016}, etc. Principally, the cavity restricts the field modes with which the matter inside the cavity can interact and allows the emitted light corresponding to those modes to be detected as it leaks out of the cavity. The light emitted from the cavity carries information about the quantum state of the system inside the cavity and can possess interesting, useful quantum properties.

Placing Rydberg atoms inside an optical cavity combines the fields of Rydberg atoms and CQED. By utilizing electromagnetically induced transparency (EIT) \cite{fleischhauer2005electromagnetically,mohapatra2007coherent} and Rydberg atom interactions, such a composite system can be very useful for both fundamental physics and applications, such as the synthesis of novel quantum states using Rydberg atom blockade that are difficult to do using other means, e.g. multi-atom entangled states. Recently, a few experiments have investigated the Rydberg atom-cavity system in both the dispersive \cite{parigi2012observation} and resonant \cite{boddeda2016rydberg,ningyuan2016observation} regimes. Intracavity EIT phenomena \cite{2007vacuum,wu2008observation,mucke2010electromagnetically,albert2011cavity,Simon2017} with Rydberg atoms has been observed in low \cite{boddeda2016rydberg} and intermediate \cite{ningyuan2016observation} finesse optical cavities.

\begin{figure}[ht]
\includegraphics[width=1\linewidth]{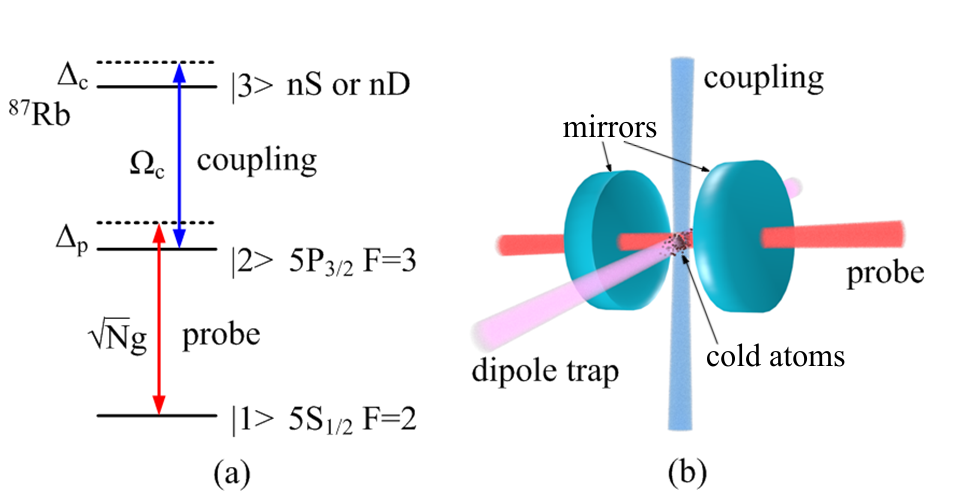}
\caption{(color online) (a) The energy level diagram of $^{87}$Rb atoms used for Rydberg EIT. (b) Schematic of the experimental setup. The atoms are transported into the high-finesse cavity from a MOT using a single beam optical dipole trap controlled by a focus-tunable lens. }\label{fig1}
\end{figure}

In this work, we report our experimental investigation of cavity assisted Rydberg EIT inside a high finesse optical cavity, $F \sim 28000$. The advantage of using a high finesse cavity is that, in principle, few-body problems can be studied inside the cavity \cite{Buchler14,Buchler16}. By using a relatively high finesse optical cavity and a small cavity size, the single atom coupling constant, $g$, can be comparable to the atomic and cavity decays. Therefore, the so-called strong-coupling regime can be achieved with only a few atoms inside the cavity. Using Rydberg atom blockade to create collective `super atoms', the atom-cavity coupling can also be increased by a factor of $\sqrt N$, where $N$ is the number of participating atoms, which can enhance the single photon emission rate for a deterministic single photon source, or even a multiple photon source \cite{kumar2016collective}.

The challenge of using a small cavity for Rydberg atoms is that when the cavity mirrors are close to the atoms, electric fields from adsorbates that stick to the cavity mirrors can significantly shift the energy levels of the Rydberg atoms \cite{sedlacek2016electric}. In this work we characterize these electric fields using the Rydberg atom cavity assisted EIT signal. We achieve a coherence time of $7.26 \pm 0.06 \,\mu$s for the system. The coherence time is a lower bound because this is approximately the dephasing time for our lasers. We demonstrate anti-crossing behavior in the atom-cavity system in the non-interacting Rydberg atom regime to show that coherent dynamics are possible in a high finesse cavity. The demonstrated coherence time is long enough to carry-out research in many of the areas mentioned earlier in the introduction.

\section{THEORETICAL ANALYSIS}
We consider a composite atom-cavity system that consists of a single-mode cavity containing $N$ three-level $^{87}$Rb atoms with a ground state $|1\rangle$ (5S$_{1/2}$, F=2), an intermediate state $|2\rangle$ (5P$_{3/2}$, F=3), and a highly excited Rydberg state $|3\rangle$ (nS or nD) driven via a two-photon transition in a ladder-type EIT configuration, as shown in Fig.~\ref{fig1}(a). The cavity mode couples the atomic transition $|1\rangle \rightarrow |2\rangle$ with a single-atom coupling constant $g = \mu \sqrt {\omega_{p} /2\hbar \varepsilon _0 V}$. Here, $\omega_p$ is the EIT probe laser frequency because we are concerned with the probe laser field that is coupled into the cavity. $\mu$ is the atomic transition dipole moment for the probe transition, $|1\rangle \rightarrow |2\rangle$, and $V$ is the cavity mode volume. The classical coupling laser drives the atomic transition $|2\rangle \rightarrow |3\rangle$ with Rabi frequency $\Omega _c$. $\Delta _p  = \omega _p  - \omega _{12}$ is the probe laser detuning and $\Delta _c  = \omega _c  - \omega _{23}$ is the coupling laser detuning. $\omega_{ij}$ is the transition frequency of the respective transitions shown in Fig.~\ref{fig1}. $\omega_c$ is the coupling laser frequency. The interaction Hamiltonian for the atom-cavity system without considering the interaction between Rydberg atoms is
\begin{equation}\label{H}
H_{{\mathop{\rm int}} }  =  - \hbar \sum\nolimits_j^N {(g\hat a\hat \sigma _{21}^{(j)} }  + \Omega _c \hat \sigma _{32}^{(j)} ) + H.c.,
\end{equation}
where $\hat \sigma _{lm}^{(j)}$ (l,m = 1,2,3) is the atomic operator for the $j^{th}$ atom and $\hat a$ is the annihilation operator for intracavity photons. The Hamiltonians for the atoms and the field are $H_{atom}  = \hbar \sum\nolimits_j^N {[\Delta _p \hat \sigma _{22}^{(j)} }  + (\Delta _p  + \Delta _c )\hat \sigma _{33}^{(j)} ]$ and $H_{field}  = \hbar \Delta _\theta  a^\dag  a$, respectively. $\Delta _\theta   = \omega _{cav}  - \omega _{12}$ is the cavity field detuning. $\omega_{cav}$ is the cavity mode frequency. For this system, the equation of motion for the intracavity field is \cite{walls2007quantum}
\begin{equation}\label{a1}
\begin{array}{l}
 \dot{\hat a} =  - \frac{i}{\hbar }[\hat a,H_{atom}  + H_{field}  + H_{{\mathop{\rm int}} } ] - \frac{{\gamma_1  + \gamma_2 }}{2}\hat a + \sqrt {\gamma_1 } \hat a_{in}  \\
 \;\;\, =  - (\frac{{\gamma_1  + \gamma_2 }}{2} - i\Delta_\theta  )\hat a + i\sum\nolimits_j^N {g\hat \sigma _{12}^{(j)} }  + \sqrt {\gamma_1 } \hat a_{in} . \\
 \end{array}
 \end{equation}
Here, $\gamma_{1,2} $ are related to the coupling constants for the external and internal cavity fields on each of the two cavity mirrors, and $\gamma = \gamma_1  = \gamma_2 $ for a symmetric cavity. In Eqn.~2 the input mirror is denoted by the subscript 1 while the output mirror is denoted by subscript 2. The transmission of each mirror is related to $\gamma$ through the round trip time of the cavity, $T_i = 2 L \gamma_i/c$. $L$ is the cavity length. See Ref.~\cite{walls2007quantum} for a more detailed discussion. The steady-state solution for the intracavity field in the frequency domain is
\begin{equation}\label{a2}
\hat a (\omega_p) = \frac{{\sqrt \gamma \hat a_{in}(\omega_p) }}{{\gamma - i\Delta   - i \frac{\omega_{p} l}{2 L}\chi }}.
\end{equation}
Here, we have assumed a symmetric cavity for simplicity of notation. $l$ is the length of the atomic sample. $\Delta = \Delta_p - \Delta_\theta = \omega_p - \omega_{cav}$ is the detuning of the input field from the cavity resonance. $\chi$ is the atomic susceptibility \cite{Min1995},
\begin{equation}\label{chi}
 \chi = \frac{{i|\mu|^2 \rho_0}}{\hbar \epsilon_0({\gamma _{12} - i\Delta_p + \frac{{|\Omega _c|^2/4 }}{{\gamma _{13} - i(\Delta _p+\Delta _c) }}})}.
\end{equation}
$\gamma_{12} = (\Gamma_1 + \Gamma_2)/2$ is the decay rate of the intermediate state, where $\Gamma_1$ is the decay rate of $|1\rangle$ and $\Gamma_2$ is the decay rate of $|2\rangle$.  $\Gamma_1$ is assumed to be zero for our calculations. $\rho_0$ is the atomic density. $\gamma_{13}= (\Gamma_1 + \Gamma_3)/2 \approx \Gamma_3/2 \approx \Gamma_{32}/2$ is the decay rate from the Rydberg state with $\Gamma_3$ the Rydberg state decay rate which is approximately the decay rate of the Rydberg state to $|5P_{3/2} \rangle$. The latter assumption is well justified because this is the largest transition frequency from the Rydberg state for any allowed transition in the physical system.  $\chi$ in Eqn.~4 is derived assuming the rotating wave approximation, a weak probe field and little population in the intermediate and Rydberg states \cite{Min1995}. The probe field  transmitted through the cavity is $|\hat{a}_{out} |^2  = \gamma |\hat{a}|^2$.

\begin{figure*}[ht]
\includegraphics[width=\linewidth]{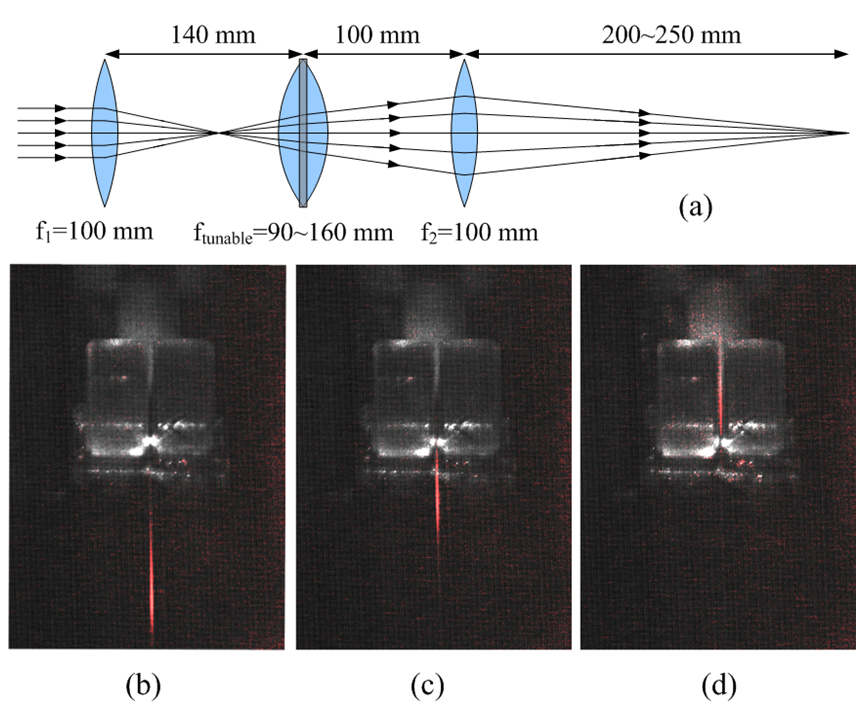}
\caption{(color online) (a) Setup of the lens system for the dipole trap. The middle lens is a focus-tunable lens, which has a focal tuning range of $90 - 160\,$mm, controlled by an applied current. (b-d)  Fluorescence images that show the process of the atoms moving into the cavity. }\label{fig2}
\end{figure*}

Under conditions of high mirror reflectivity, $R\approx 1$, and a small round trip phase shift, these results are equivalent to a semi-classical formula for the cavity transmission function where the linear dispersive and absorptive properties of the intracavity medium are taken into account \cite{sheng2011understanding}.
The intensity transmission function of the coupled atom-cavity system in this picture is \cite{sheng2011understanding}
\begin{equation}\label{S}
S(\omega _p ) = \frac{{T^2 }}{{1 + R^2 \alpha ^2  - 2R\alpha \cos [(\Delta  + (\omega _p {l/2L)}\chi ')2L/c]}},
\end{equation}
where $\alpha  \equiv \exp [ - \omega _p l\chi'' /c]$ describes the intracavity absorption, $R$ is the reflectivity of the mirrors and $\chi  = \chi ' + i\chi''$ with $\chi '$ being the dispersive and $\chi''$ the absorptive components of the susceptibility. If $R \approx 1$ and both the absorption and cosine term are expanded to first order, Eqn.~5 is proportional to
\begin{equation}
\frac{\mid \hat a_{out} \mid^2}{\mid \hat a_{in} \mid^2} = \frac{T^2}{|T - i \frac{2L}{c}(\Delta + \frac{\omega_p l \chi}{2 L})|^2},
\end{equation}
which can be obtained from Eqn.~3 and $|\hat{a}_{out} |^2  = \gamma |\hat{a}|^2$. We use Eqn.~5 to analyze our data.

When the argument of the cosine term of Eqn.~5 is an integral of $\pi$, the transmission through the cavity is a maximum. The peaks in the spectrum are determined by the condition $\Delta = -\omega _p \rm{(l/2L)}\chi\rm{'}$. The cavity transmission, then, measures the phase shifts caused by the light-matter interactions. Theoretically, there are five transmission peaks in the cavity assisted EIT spectrum that correspond to conditions where strong light-matter interactions occur. The central peak results from the normal dispersion associated with the preparation of the EIT dark state. The two pairs of symmetrically distributed side peaks that are observed in the cavity assisted EIT transmission are associated with states that can absorb light. The two peaks lying closest to central EIT dark state are the result of the dispersion where the slope is negative, or anomalous dispersion. These features of the spectrum are difficult to detect because the absorption is large at these detunings. The outer, bright state peaks appear on the wings of the atomic absorption peak so they are more readily observed. These two outer peaks correspond to normal dispersion. The phase shifts that minimize the cosine term in the cavity transmission are an indication of strong light-matter interaction so they can be identified as polaritons. The `bright' polaritons are associated with the absorbing matter states and the `dark' polariton is associated with non-absorbing matter states.

\begin{figure}[ht]
\includegraphics[width=1\linewidth]{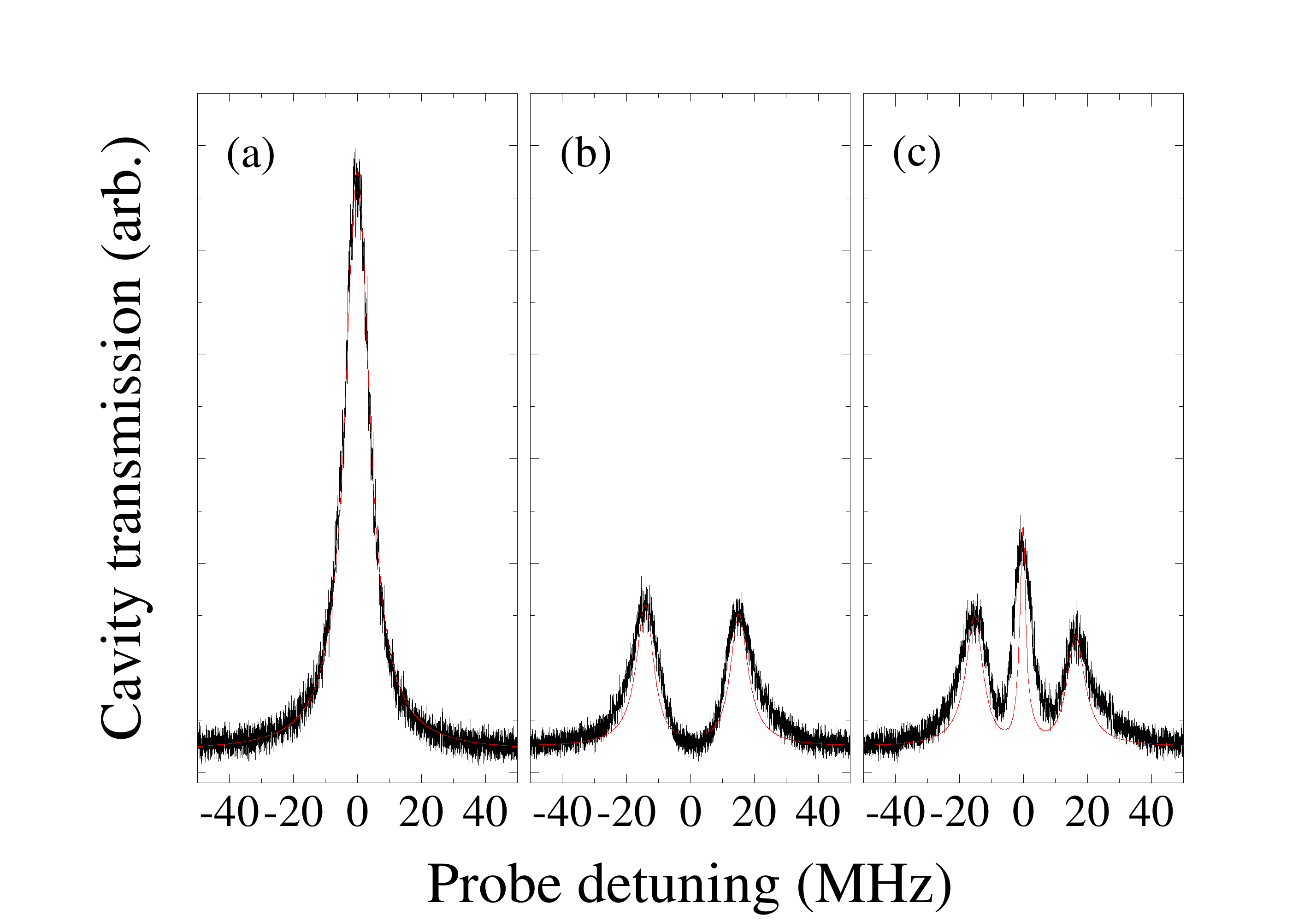}
\caption{(color online) Measured cavity transmission versus probe laser detuning. The thick black line is the experimental data. The thin red line is the theory corresponding to Eqn.~5 of the text. The data is the average of 30 traces. $\Omega_p/2\pi = 9.1\,$MHz. (a) Empty cavity. (b) Two-level atoms, i.e. $\Omega _c  = 0\,$MHz. (c) Rydberg EIT for the 35S$_{1/2}$ Rydberg state, $\Omega _c /2\pi  = 4.1\,$MHz. There are $\sim 25$ atoms in the interaction region in the cavity. There is some instrumental broadening in these traces caused by the averaging and the signal integration.}\label{fig3}
\end{figure}

\section{EXPERIMENTAL}
The experiment is performed by exciting ultracold $^{87}$Rb atoms to Rydberg states inside an optical cavity, as shown in Fig.~\ref{fig1}(b). The atoms are first loaded into a single beam optical dipole trap at $1064\,$nm with a calculated $420\,\mu$K depth from a magneto-optical trap (MOT), which is $\sim 2\,$cm away from the cavity. We estimate the temperature of the atoms in the trap to be $< 10 \,\mu$K. The atoms are transported into the cavity using a focus-tunable lens (Optotune EL-10-30-Ci). By properly designing the optical system, as shown in Fig.~\ref{fig2}(a), the focal position of the dipole trap laser can be dynamically controlled while maintaining a constant beam waist, which minimizes the loss and heating of the atoms in the trap during the transportation \cite{Esslinger14}. The beam waist is $\sim 30 \,\mu$m. We use $\sim 4\,$W of light to form the atom trap.  Fig.~\ref{fig2}(b)-(d) show the fluorescence images of moving atoms as they are transported to the cavity. The Rydberg EIT coupling laser light, $\sim 480\,$nm, is injected through the gap between the cavity mirrors. The coupling laser beam spot size is $\sim 30\,\mu$m. The probe laser, $\sim 780\,$nm, propagates along the cavity axis. The light transmitted through the cavity is coupled into an optical fiber to minimize the background noise and detected by a photomultiplier tube detector (PMT) (Hamamatsu H10721-20).

The cavity consists of two $7.75\,$mm diameter mirrors with $25\,$mm radii of curvature separated by $900\,\mu$m. The reflectivity of the input mirror is $99.9985 \%$ and the reflectivity of the output mirror is $99.985 \%$. The measured finesse of the cavity is $F \sim 28000$. The waist size of the TEM$_{00}$ mode of the cavity is $\sim 30\,\mu$m. The relevant CQED parameters for the system are $(g ,\kappa ,\gamma_{12} )/2\pi  = (3\,\textrm{MHz},3\,\textrm{MHz},3\,\textrm{MHz})$, where $\kappa$ and $\gamma_{12}$ are the half-width-half-maximum of the cavity and atomic decays, respectively. The corresponding cooperativity is $C = 0.125$. To control the length of the cavity, the mirrors are mounted on two shear-mode piezoelectric transducers (PZT) (Noliac CSAP02), which are attached to a copper block. The PZTs are topped by a grounded, aluminum foil sheet in order to shield the atoms from the electric field created by the voltages applied to the PZTs. The copper block sits on RTV silicone (Dow Corning 736) to provide vibration isolation. A heater (Watlow CER-1-01-00335) is attached to the copper block to heat the block above room temperature. The cavity is typically heated to $\sim 50^{\circ}$C. The temperature is regulated with an external temperature control circuit. The cavity is actively stabilized by locking it to a transfer laser at $852\,$nm, which is frequency stabilized by locking it to an ultra-stable cavity. The cavity frequency can be tuned by changing the frequency of the sideband used to lock the transfer laser \cite{Milani17}.

\begin{figure}[ht]
\includegraphics[width=1\linewidth]{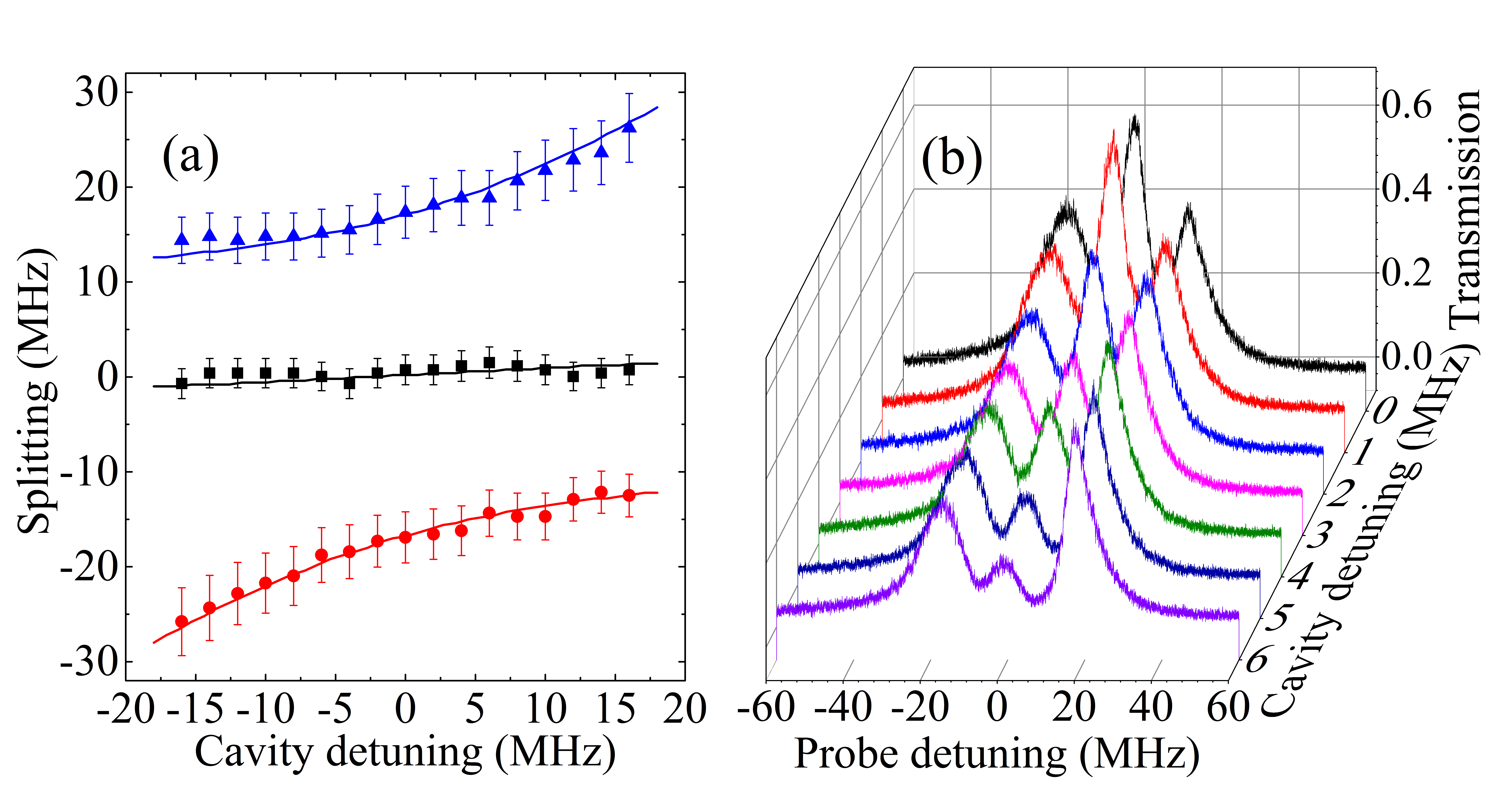}
\caption{(color online) (a) The positions of the cavity transmission peaks as a function of cavity detuning, $\Delta_\theta /2\pi$, for the $35S_{1/2}$ Rydberg state. The solid lines are the peak centers calculated from Eqn.~5. (b) The cavity transmission spectra for different cavity detunings as a function of probe laser detuning. The parameters are the same as in Fig.~\ref{fig3}.}\label{fig4}
\end{figure}

\section{DISCUSSION}
Figure~\ref{fig3} shows the measured cavity transmission of the probe laser versus the probe detuning, $\Delta_p /2\pi$. The black lines are experimental data and the red lines are theoretical calculations based on Eqn.~5. The transmission peak of the empty cavity, i.e. no atoms in the cavity, is plotted in Fig.~\ref{fig3}(a) as a reference. The probe laser frequency is scanned by an AOM in a double-pass configuration and a noise-eater is used to stabilize the probe laser power.  Fig.~\ref{fig3}(b) shows the probe laser cavity transmission when two-level atoms are inside the cavity, i.e. only the probe field is applied. We observe two transmission peaks, the normal atom-cavity modes, at $\Delta  =  \pm \sqrt N g$, from which we estimate that there are $\sim$ 25 atoms in the interaction region which has dimensions $\sim 30 \times 30 \times 30\,\mu m^3$ (not the cavity mode volume). The atomic density in the optical dipole trap, or the atom number, can be controlled by changing the position where the dipole trap is loaded relative to the MOT, which causes a change of the atomic density of more than two orders of magnitude. The change can be observed by measuring the frequency splitting shown in Fig.~\ref{fig3}(b).

The optical dipole trap is on during the measurements, which results in an effective AC Stark shift of the EIT resonance conditions of $\sim 16\,$MHz. The atoms at different locations inside the trap can experience different detunings due to the AC Stark shift. These spatially dependent energy shifts can lead to inhomogeneous broadening of the cavity transmission spectrum. Inhomogeneous broadening can hinder the observation of the intracavity EIT spectrum and affect the measured lineshape.

\begin{figure}[ht]
\includegraphics[width=1\linewidth]{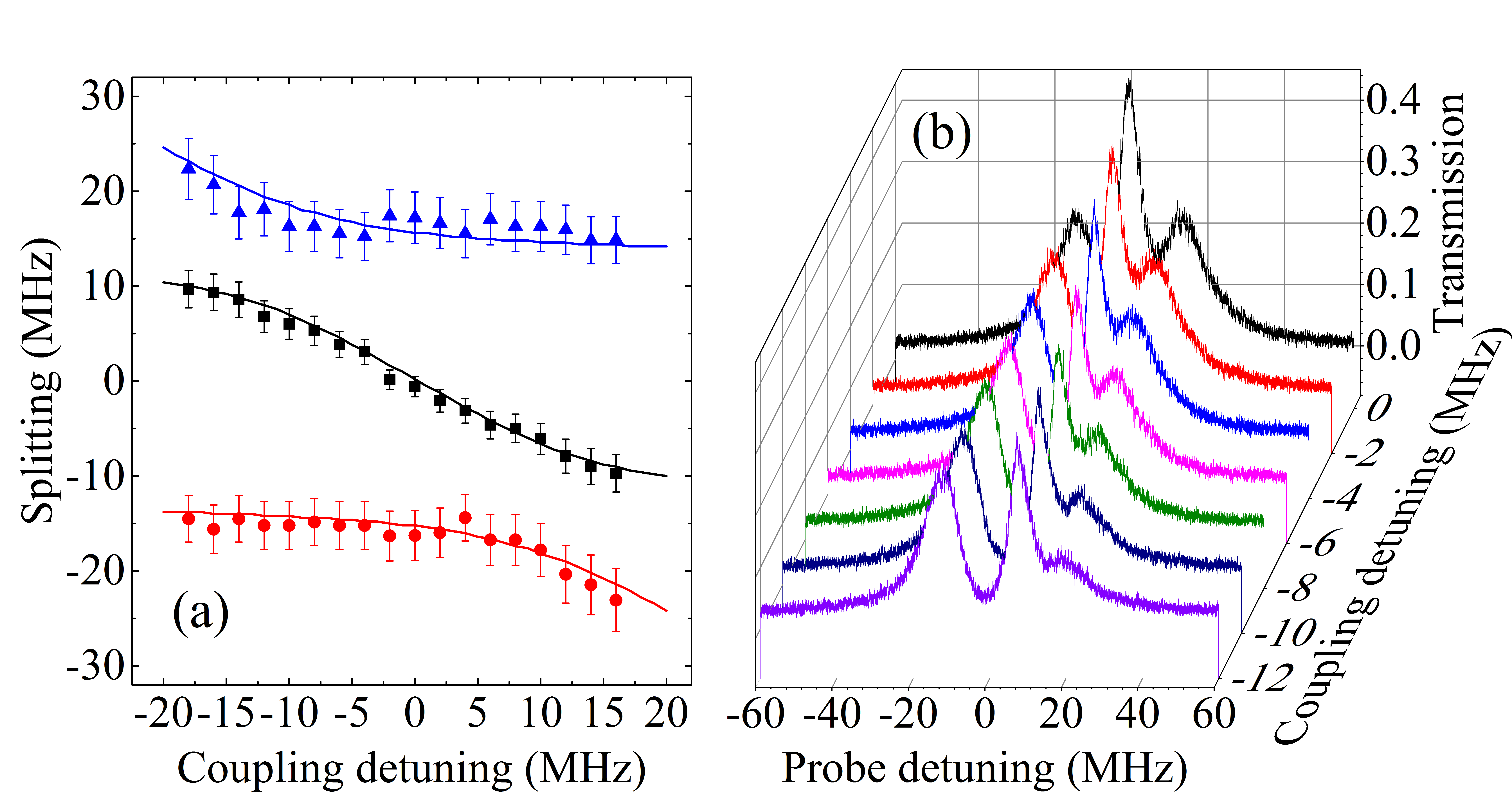}
\caption{(color online)(a) The positions of the cavity transmission peaks as a function of coupling laser detuning, $\Delta _c /2\pi$, for the $35S_{1/2}$ Rydberg state. The solid lines are the peak centers calculated from Eqn.~5. (b) The cavity transmission spectra for different coupling detunings as a function of probe laser detuning. The parameters are the same as in Fig.~\ref{fig3}.}\label{fig5}
\end{figure}

When the Rydberg coupling laser is turned on, the cavity transmission spectrum exhibits a three peak structure, as shown in Fig.~\ref{fig3}(c). The two side peaks, shown detuned from the frequency zero on the trace, are caused by bright-state polaritons as in Fig.~\ref{fig3}(b). The central peak at zero frequency corresponds to a dark-state polariton. A relatively low-lying Rydberg state, $35S_{1/2}$, is used in Fig.~\ref{fig3}(c), so that interactions between Rydberg atoms can be ignored. Moreover, the coupling Rabi frequency is larger for a fixed blue laser power and the DC-Stark shift from stray electric fields is smaller as the Rydberg principal quantum number, $n$, decreases. Despite the trapping field, cavity assisted EIT is observed.

To further verify that cavity EIT is possible despite the spatially dependent AC Stark shifts, a lambda-type EIT configuration was first used to measure the inhomogeneous broadening. Using the $5S_{1/2}(F=2)\rightarrow 5P_{3/2}(F=2)\rightarrow 5S_{1/2}(F=1)$ EIT system we measured an EIT cavity transmission linewidth of $1\,$MHz. A laser locked to a saturated absorption setup was used for the coupling laser. The linewidth is limited by the coupling laser linewidth, $\sim 1\,$MHz. The transmission linewidth indicates that the inhomogeneous AC Stark shift across the sample interacting with the cavity and lasers is not significant enough to change the three-level EIT dispersion curve to an effective two-level dispersion curve \cite{hui2014cavity}, supporting the data shown in Fig.~\ref{fig3}(c). The result is consistent with the fact that the atoms sit in the bottom of the dipole trap since they are at a temperature of $\leqslant 10\,\mu$K while the trap depth is $\sim 420\,\mu$K. In principle, the broadening can be reduced by turning off the trap for the measurements or by further cooling the sample, but later in the paper, when we address static electric and magnetic fields, we will show that linewidths of $\sim 140\,$kHz can be obtained, further supporting these initial observations.

We observed anti-crossing behavior in both the cavity field detuning and coupling laser detuning cases for the $35S_{1/2}$ Rydberg state. In Fig.~\ref{fig4}(a), we plot the positions of the three cavity transmission peaks as a function of the cavity field detuning, $\Delta_\theta /2\pi$ with the probe laser on resonance. The anti-crossing behavior shown in the plot is straightforward to explain, as in coupled two-level atom-cavity systems \cite{gripp1997evolution}, it arises from the mixing of matter and field oscillations. The center peak position is not very sensitive to the cavity detuning due to the steep dispersion near the EIT resonance. The slope of the dispersion at the center peak in Fig.~\ref{fig4}(a) is approximately inversely proportional to the frequency-locking coefficient  $\eta \sim \partial \chi '/\partial \omega _p$ \cite{lukin1998intracavity,wang2000cavity}. Fig.~\ref{fig4}(b) shows the probe laser transmission spectrum as a function of cavity detuning.

 Fig.~\ref{fig5} shows the measured cavity transmission spectra for different coupling laser detunings, $\Delta_c /2\pi$. Similar to the cavity field detuning case, the anti-crossing behavior for the two side peaks can clearly be observed in the coupling laser detuning case, Fig.~\ref{fig5}(a). The cavity transmission spectra for different coupling laser detunings are presented in Fig.~\ref{fig5}(b).

 The observations in Fig.~\ref{fig4} and \ref{fig5} demonstrate that it is possible to observe cavity assisted Rydberg EIT in a high-finesse cavity. Electric fields do not change the dispersion enough to eliminate the cavity assisted Rydberg EIT signal. The observation of cavity assisted Rydberg EIT here leads to the question: what is the coherence time that can be achieved in the high-finesse cavity?

\begin{figure}[ht]
\includegraphics[width=0.65\linewidth]{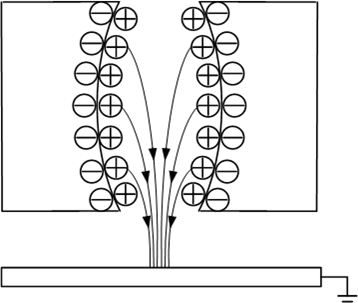}
\caption{This figure shows a sketch of the electric field inside our cavity as described in the text.}\label{fieldsketch}
\end{figure}

Electric field gradients can cause dephasing that will limit the coherence time. During the experiment, we observed a relative large electric field near the center of the cavity, $ 1.5 \pm 0.1\,$V$\,$cm$^{-1}$. The value of the electric field was determined by measuring Stark splittings of the $38D_{5/2}$ and $33D_{3/2}$ Rydberg states inside the cavity. In the course of these measurements and others with the $35S_{1/2}$ Rydberg state, we also observed a magnetic field of $1\pm 0.3\,$G. The magnetic field was more difficult to determine precisely since there were energy shifts from both the $5S_{1/2}$ ground state and the Rydberg states. We exclude the possibility that the electric fields are generated by the PZTs or the heater. We operated the system at different heater currents and different PZT voltages and observed the same Stark shifts and line shapes. We conclude that the electric field is due to adsorbates on the surfaces of the mirrors. Rb can deposit onto the mirrors and polarize as it interacts with the surface. These small dipoles when summed over a surface can produce a macroscopic electric field. A sketch of this effect in our cavity setup is shown in Fig.~\ref{fieldsketch}. In a recent experimental study, we arrived at a similar conclusion after a detailed study of the adsorbate electric field caused by Rb atoms on a quartz surface \cite{sedlacek2016electric}. In support of this assertion, we were able to change the size of the electric field when the EIT coupling laser beam was misaligned so that it was scattering off one of the cavity mirrors. Presumably, the light was desorbing Rb from the mirror surfaces causing a change in the electric field \cite{Arlt06}. Note that the scattering of the $480\,$nm light can increase the electric field because the dipole density on the mirrors can become imbalanced. We also used ultra-violet light to change the electric field inside the cavity further supporting the idea that the surface adsorbates are the primary source of the electric field. We were able to reduce the electric field by as much as $1\,$V$\,$cm$^{-1}$ using ultra-violet light generated by an array of light emitting diodes. We did not observe any effects associated with charging of the mirror surface when the LEDs were used.

\begin{figure}[ht]
\includegraphics[width=1\linewidth]{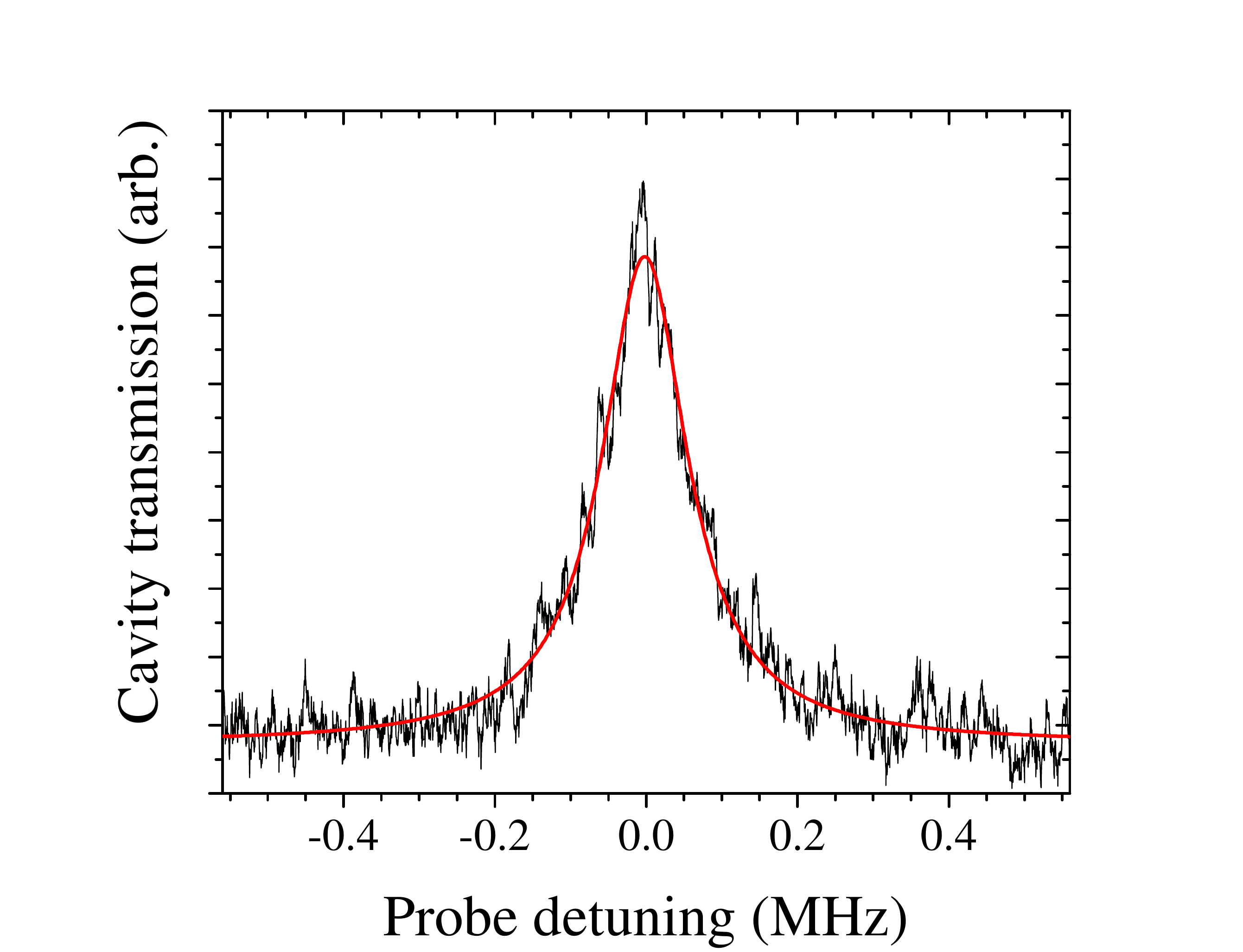}
\caption{(color online) This figure shows a cavity transmission peak corresponding to a dark state polariton formed on the $5S_{1/2}(m_F=0) \rightarrow 5P_{3/2} \rightarrow 35S_{1/2}(m_F=0)$ transition. The linewidth is $138 \pm 1$kHz. The linewidth corresponds to a coherence time of $7.26 \pm 0.06 \,\mu$s. The data was taken using the $35S_{1/2}$ Rydberg state. The probe and coupling Rabi frequencies were $\Omega_p/2 \pi = 1.2\,$MHz and $\Omega_c/2 \pi = 5.2\,$MHz, respectively. There were approximately $25$ atoms in the interaction region of the cavity. The red solid line is a fit to a Lorentzian. The black solid line is the experimental data. A magnetic field of $\sim 3.6\,$G was applied with a single coil and used for the measurement. The atomic sample was located on the symmetry axis of the coil.}\label{coherencetime}
\end{figure}

Although the electric field is significant for a Rydberg atom, a constant energy shift is not necessarily relevant to many applications. On the other hand, electric field gradients are important because they can shorten the coherence time of the system. To investigate the coherence time of the system we split the Rydberg state with a magnetic field to isolate transitions corresponding to specific $m_F$ transitions. Fig.~\ref{coherencetime} shows a cavity EIT transmission signal as a function of $\Delta_p/2 \pi$ corresponding to a coherence time of $7.26 \pm 0.06 \,\mu$s. The graph is constructed by averaging $36$ probe laser sweeps. The linewidth of the spectrum is determined predominantly by the spectral widths of the probe and coupling lasers, since the convolution of the spectral width of each of the two lasers derived from their locking signals is $\sim 130\,$kHz. The expected broadening due to the AC Stark shifts in the trap is similar in magnitude to the broadening caused by the lasers. It may be possible to improve the coherence time by improving the locking of the probe and coupling lasers as well as turning off the dipole trap during the experiment, although Doppler shifts, the natural linewidth of the Rydberg state and collisions will limit the coherence time at some point \cite{kumar2017}, depending on the specific experimental parameters used.

The fact that electric field gradients play only a small role in decohering the system results from the small interaction volume in the cavity and the fact that the adsorbates on the surface cover the surfaces uniformly. The mirrors are highly polished and the dielectric coatings are high reflectivity so there is no reason to believe that certain areas of the mirror near the trap are coated more than other areas, since the adsorbate coatings are determined by the thermodynamics and the adsorbate binding energy \cite{sedlacek2016electric}. At $50^\circ\,$C, the coating is less than a monolayer for similar materials. The precise binding energy of a Rb atom adsorbed to the dielectric stack on the polished substrate is not known, but the outer protective oxide layers are similar to quartz which would have a fraction of a monolayer of Rb at these temperatures and pressures, dominated by the atoms used to load the cavity \cite{sedlacek2016electric}. The Rb binds to oxygen atoms at the surface in the case of quartz and we expect similar behavior here. There is not enough Rb on the mirrors to significantly change the finesse of the cavity because we have not observed any changes associated with the cavity that correlate with the intracavity electric fields.

In conclusion, we observed dark state polaritons via cavity assisted EIT in a high finesse cavity ($\sim 28000$) and characterized them as a function of cavity and coupling detuning. Perhaps more interestingly, we studied the electric fields generated by adsorbates adhering to the cavity mirrors. We were able to put a lower limit on the system coherence time of $7.26 \pm 0.06 \,\mu$s. The coherence time that we measured is limited primarily by our laser coherence times. The measured coherence time is long enough to do many interesting experiments, such as studying collective excitations of Rydberg blockaded samples in the low excitation limit in the cavity. We believe that this work helps to open the way for experiments on wavefunction manipulation with Rydberg blockade by enabling its study through the analysis of the photons produced by these interesting entangled states. Achieving the strong coupling regime with a collective Rydberg interaction is possible to do by using the $\sqrt{N}$ enhancement to the coupling, reducing the cavity mode volume further, or some combination of the two.

\section{Acknowledgments}
This work was supported by the AFOSR (FA9550-12-1-0282) and NSF (PHY-1104424).

\bibliographystyle{apsrev4-1}
%\bibliography{ref}

\begin{thebibliography}{61}%
\makeatletter
\providecommand \@ifxundefined [1]{%
 \@ifx{#1\undefined}
}%
\providecommand \@ifnum [1]{%
 \ifnum #1\expandafter \@firstoftwo
 \else \expandafter \@secondoftwo
 \fi
}%
\providecommand \@ifx [1]{%
 \ifx #1\expandafter \@firstoftwo
 \else \expandafter \@secondoftwo
 \fi
}%
\providecommand \natexlab [1]{#1}%
\providecommand \enquote  [1]{``#1''}%
\providecommand \bibnamefont  [1]{#1}%
\providecommand \bibfnamefont [1]{#1}%
\providecommand \citenamefont [1]{#1}%
\providecommand \href@noop [0]{\@secondoftwo}%
\providecommand \href [0]{\begingroup \@sanitize@url \@href}%
\providecommand \@href[1]{\@@startlink{#1}\@@href}%
\providecommand \@@href[1]{\endgroup#1\@@endlink}%
\providecommand \@sanitize@url [0]{\catcode `\\12\catcode `\$12\catcode
  `\&12\catcode `\#12\catcode `\^12\catcode `\_12\catcode `\%12\relax}%
\providecommand \@@startlink[1]{}%
\providecommand \@@endlink[0]{}%
\providecommand \url  [0]{\begingroup\@sanitize@url \@url }%
\providecommand \@url [1]{\endgroup\@href {#1}{\urlprefix }}%
\providecommand \urlprefix  [0]{URL }%
\providecommand \Eprint [0]{\href }%
\providecommand \doibase [0]{http://dx.doi.org/}%
\providecommand \selectlanguage [0]{\@gobble}%
\providecommand \bibinfo  [0]{\@secondoftwo}%
\providecommand \bibfield  [0]{\@secondoftwo}%
\providecommand \translation [1]{[#1]}%
\providecommand \BibitemOpen [0]{}%
\providecommand \bibitemStop [0]{}%
\providecommand \bibitemNoStop [0]{.\EOS\space}%
\providecommand \EOS [0]{\spacefactor3000\relax}%
\providecommand \BibitemShut  [1]{\csname bibitem#1\endcsname}%
\let\auto@bib@innerbib\@empty
%</preamble>
\bibitem [{\citenamefont {Gallagher}(2005)}]{gallagher2005rydberg}%
  \BibitemOpen
  \bibfield  {author} {\bibinfo {author} {\bibfnamefont {T.~F.}\ \bibnamefont
  {Gallagher}},\ }\href@noop {} {\emph {\bibinfo {title} {Rydberg atoms}}}\
  (\bibinfo  {publisher} {Cambridge University Press},\ \bibinfo {year}
  {2005})\BibitemShut {NoStop}%
\bibitem [{\citenamefont {Tong}\ \emph {et~al.}(2004)\citenamefont {Tong},
  \citenamefont {Farooqi}, \citenamefont {Stanojevic}, \citenamefont
  {Krishnan}, \citenamefont {Zhang}, \citenamefont {Cote}, \citenamefont
  {Eyler},\ and\ \citenamefont {Gould}}]{tong2004local}%
  \BibitemOpen
  \bibfield  {author} {\bibinfo {author} {\bibfnamefont {D.}~\bibnamefont
  {Tong}}, \bibinfo {author} {\bibfnamefont {S.M.}~\bibnamefont {Farooqi}},
  \bibinfo {author} {\bibfnamefont {J.}~\bibnamefont {Stanojevic}}, \bibinfo
  {author} {\bibfnamefont {S.}~\bibnamefont {Krishnan}}, \bibinfo {author}
  {\bibfnamefont {Y.P.}~\bibnamefont {Zhang}}, \bibinfo {author} {\bibfnamefont
  {R.}~\bibnamefont {Cote}}, \bibinfo {author} {\bibfnamefont {E.E.}~\bibnamefont
  {Eyler}}, \ and\ \bibinfo {author} {\bibfnamefont {P.L.}~\bibnamefont
  {Gould}},\ }\href@noop {} {\bibfield  {journal} {\bibinfo  {journal} {Phys.
  Rev. Lett.}\ }\textbf {\bibinfo {volume} {93}},\ \bibinfo {pages} {063001}
  (\bibinfo {year} {2004})}\BibitemShut {NoStop}%
\bibitem [{\citenamefont {Heidemann}\ \emph {et~al.}(2007)\citenamefont
  {Heidemann}, \citenamefont {Raitzsch}, \citenamefont {Bendkowsky},
  \citenamefont {Butscher}, \citenamefont {Low}, \citenamefont {Santos},\ and\
  \citenamefont {Pfau}}]{heidemann2007evidence}%
  \BibitemOpen
  \bibfield  {author} {\bibinfo {author} {\bibfnamefont {R.}~\bibnamefont
  {Heidemann}}, \bibinfo {author} {\bibfnamefont {U.}~\bibnamefont {Raitzsch}},
  \bibinfo {author} {\bibfnamefont {V.}~\bibnamefont {Bendkowsky}}, \bibinfo
  {author} {\bibfnamefont {B.}~\bibnamefont {Butscher}}, \bibinfo {author}
  {\bibfnamefont {R.}~\bibnamefont {Low}}, \bibinfo {author} {\bibfnamefont
  {L.}~\bibnamefont {Santos}}, \ and\ \bibinfo {author} {\bibfnamefont
  {T.}~\bibnamefont {Pfau}},\ }\href@noop {} {\bibfield  {journal} {\bibinfo
  {journal} {Phys. Rev. Lett.}\ }\textbf {\bibinfo {volume} {99}},\ \bibinfo
  {pages} {163601} (\bibinfo {year} {2007})}\BibitemShut {NoStop}%
\bibitem [{\citenamefont {Comparat}\ and\ \citenamefont
  {Pillet}(2010)}]{comparat2010dipole}%
  \BibitemOpen
  \bibfield  {author} {\bibinfo {author} {\bibfnamefont {D.}~\bibnamefont
  {Comparat}}\ and\ \bibinfo {author} {\bibfnamefont {P.}~\bibnamefont
  {Pillet}},\ }\href@noop {} {\bibfield  {journal} {\bibinfo  {journal} {JOSA
  B}\ }\textbf {\bibinfo {volume} {27}},\ \bibinfo {pages} {A208} (\bibinfo
  {year} {2010})}\BibitemShut {NoStop}%
\bibitem [{\citenamefont {Marcassa}\ and\ \citenamefont
  {Shaffer}(2014)}]{marcassa2014interactions}%
  \BibitemOpen
  \bibfield  {author} {\bibinfo {author} {\bibfnamefont {L.~G.}\ \bibnamefont
  {Marcassa}}\ and\ \bibinfo {author} {\bibfnamefont {J.~P.}\ \bibnamefont
  {Shaffer}},\ }\href@noop {} {\bibfield  {journal} {\bibinfo  {journal} {Adv.
  At. Mol. Opt. Phys.}\ }\textbf {\bibinfo {volume} {63}},\ \bibinfo {pages}
  {47} (\bibinfo {year} {2014})}\BibitemShut {NoStop}%
\bibitem [{\citenamefont {Saffman}\ \emph {et~al.}(2010)\citenamefont
  {Saffman}, \citenamefont {Walker},\ and\ \citenamefont
  {Molmer}}]{saffman2010quantum}%
  \BibitemOpen
  \bibfield  {author} {\bibinfo {author} {\bibfnamefont {M.}~\bibnamefont
  {Saffman}}, \bibinfo {author} {\bibfnamefont {T.}~\bibnamefont {Walker}}, \
  and\ \bibinfo {author} {\bibfnamefont {K.}~\bibnamefont {Molmer}},\
  }\href@noop {} {\bibfield  {journal} {\bibinfo  {journal} {Rev. Mod. Phys.}\
  }\textbf {\bibinfo {volume} {82}},\ \bibinfo {pages} {2313} (\bibinfo {year}
  {2010})}\BibitemShut {NoStop}%
\bibitem [{\citenamefont {Saffman}(2016)}]{saffman2016}%
  \BibitemOpen
  \bibfield  {author} {\bibinfo {author} {\bibfnamefont {M.}~\bibnamefont
  {Saffman}},\ }\href {http://stacks.iop.org/0953-4075/49/i=20/a=202001}
  {\bibfield  {journal} {\bibinfo  {journal} {Journal of Physics B: Atomic,
  Molecular and Optical Physics}\ }\textbf {\bibinfo {volume} {49}},\ \bibinfo
  {pages} {202001} (\bibinfo {year} {2016})}\BibitemShut {NoStop}%
\bibitem [{\citenamefont {Pohl}\ \emph {et~al.}(2009)\citenamefont {Pohl},
  \citenamefont {Sadeghpour},\ and\ \citenamefont {Schmelcher}}]{pohl2009cold}%
  \BibitemOpen
  \bibfield  {author} {\bibinfo {author} {\bibfnamefont {T.}~\bibnamefont
  {Pohl}}, \bibinfo {author} {\bibfnamefont {H.~R.}\ \bibnamefont
  {Sadeghpour}}, \ and\ \bibinfo {author} {\bibfnamefont {P.}~\bibnamefont
  {Schmelcher}},\ }\href@noop {} {\bibfield  {journal} {\bibinfo  {journal}
  {Phys. Rep.}\ }\textbf {\bibinfo {volume} {484}},\ \bibinfo {pages} {181}
  (\bibinfo {year} {2009})}\BibitemShut {NoStop}%
\bibitem [{\citenamefont {Fan}\ \emph {et~al.}(2015)\citenamefont {Fan},
  \citenamefont {Kumar}, \citenamefont {Sedlacek}, \citenamefont {Kubler},
  \citenamefont {Karimkashi},\ and\ \citenamefont {Shaffer}}]{fan2015atom}%
  \BibitemOpen
  \bibfield  {author} {\bibinfo {author} {\bibfnamefont {H.}~\bibnamefont
  {Fan}}, \bibinfo {author} {\bibfnamefont {S.}~\bibnamefont {Kumar}}, \bibinfo
  {author} {\bibfnamefont {J.}~\bibnamefont {Sedlacek}}, \bibinfo {author}
  {\bibfnamefont {H.}~\bibnamefont {Kubler}}, \bibinfo {author} {\bibfnamefont
  {S.}~\bibnamefont {Karimkashi}}, \ and\ \bibinfo {author} {\bibfnamefont
  {J.~P.}\ \bibnamefont {Shaffer}},\ }\href@noop {} {\bibfield  {journal}
  {\bibinfo  {journal} {J. Phys. B}\ }\textbf {\bibinfo {volume} {48}},\
  \bibinfo {pages} {202001} (\bibinfo {year} {2015})}\BibitemShut {NoStop}%
\bibitem [{\citenamefont {Gorshkov}\ \emph {et~al.}(2011)\citenamefont
  {Gorshkov}, \citenamefont {Otterbach}, \citenamefont {Fleischhauer},
  \citenamefont {Pohl},\ and\ \citenamefont {Lukin}}]{gorshkov2011photon}%
  \BibitemOpen
  \bibfield  {author} {\bibinfo {author} {\bibfnamefont {A.~V.}\ \bibnamefont
  {Gorshkov}}, \bibinfo {author} {\bibfnamefont {J.}~\bibnamefont {Otterbach}},
  \bibinfo {author} {\bibfnamefont {M.}~\bibnamefont {Fleischhauer}}, \bibinfo
  {author} {\bibfnamefont {T.}~\bibnamefont {Pohl}}, \ and\ \bibinfo {author}
  {\bibfnamefont {M.~D.}\ \bibnamefont {Lukin}},\ }\href@noop {} {\bibfield
  {journal} {\bibinfo  {journal} {Physical review letters}\ }\textbf {\bibinfo
  {volume} {107}},\ \bibinfo {pages} {133602} (\bibinfo {year}
  {2011})}\BibitemShut {NoStop}%
\bibitem [{\citenamefont {Dudin}\ and\ \citenamefont
  {Kuzmich}(2012)}]{dudin2012strongly}%
  \BibitemOpen
  \bibfield  {author} {\bibinfo {author} {\bibfnamefont {Y.}~\bibnamefont
  {Dudin}}\ and\ \bibinfo {author} {\bibfnamefont {A.}~\bibnamefont
  {Kuzmich}},\ }\href@noop {} {\bibfield  {journal} {\bibinfo  {journal}
  {Science}\ }\textbf {\bibinfo {volume} {336}},\ \bibinfo {pages} {887}
  (\bibinfo {year} {2012})}\BibitemShut {NoStop}%
\bibitem [{\citenamefont {Peyronel}\ \emph {et~al.}(2012)\citenamefont
  {Peyronel}, \citenamefont {Firstenberg}, \citenamefont {Liang}, \citenamefont
  {Hofferberth}, \citenamefont {Gorshkov}, \citenamefont {Pohl}, \citenamefont
  {Lukin},\ and\ \citenamefont {Vuleti{\'c}}}]{peyronel2012quantum}%
  \BibitemOpen
  \bibfield  {author} {\bibinfo {author} {\bibfnamefont {T.}~\bibnamefont
  {Peyronel}}, \bibinfo {author} {\bibfnamefont {O.}~\bibnamefont
  {Firstenberg}}, \bibinfo {author} {\bibfnamefont {Q.-Y.}\ \bibnamefont
  {Liang}}, \bibinfo {author} {\bibfnamefont {S.}~\bibnamefont {Hofferberth}},
  \bibinfo {author} {\bibfnamefont {A.~V.}\ \bibnamefont {Gorshkov}}, \bibinfo
  {author} {\bibfnamefont {T.}~\bibnamefont {Pohl}}, \bibinfo {author}
  {\bibfnamefont {M.~D.}\ \bibnamefont {Lukin}}, \ and\ \bibinfo {author}
  {\bibfnamefont {V.}~\bibnamefont {Vuleti{\'c}}},\ }\href@noop {} {\bibfield
  {journal} {\bibinfo  {journal} {Nature}\ }\textbf {\bibinfo {volume} {488}},\
  \bibinfo {pages} {57} (\bibinfo {year} {2012})}\BibitemShut {NoStop}%
\bibitem [{\citenamefont {Pritchard}\ \emph {et~al.}(2012)\citenamefont
  {Pritchard}, \citenamefont {Adams},\ and\ \citenamefont
  {M{\o}lmer}}]{pritchard2012correlated}%
  \BibitemOpen
  \bibfield  {author} {\bibinfo {author} {\bibfnamefont {J.D.}~\bibnamefont
  {Pritchard}}, \bibinfo {author} {\bibfnamefont {C.S}~\bibnamefont {Adams}}, \
  and\ \bibinfo {author} {\bibfnamefont {K.}~\bibnamefont {M{\o}lmer}},\
  }\href@noop {} {\bibfield  {journal} {\bibinfo  {journal} {Physical review
  letters}\ }\textbf {\bibinfo {volume} {108}},\ \bibinfo {pages} {043601}
  (\bibinfo {year} {2012})}\BibitemShut {NoStop}%
\bibitem [{\citenamefont {Firstenberg}\ \emph {et~al.}(2013)\citenamefont
  {Firstenberg}, \citenamefont {Peyronel}, \citenamefont {Liang}, \citenamefont
  {Gorshkov}, \citenamefont {Lukin},\ and\ \citenamefont
  {Vuleti{\'c}}}]{firstenberg2013attractive}%
  \BibitemOpen
  \bibfield  {author} {\bibinfo {author} {\bibfnamefont {O.}~\bibnamefont
  {Firstenberg}}, \bibinfo {author} {\bibfnamefont {T.}~\bibnamefont
  {Peyronel}}, \bibinfo {author} {\bibfnamefont {Q.-Y.}\ \bibnamefont {Liang}},
  \bibinfo {author} {\bibfnamefont {A.~V.}\ \bibnamefont {Gorshkov}}, \bibinfo
  {author} {\bibfnamefont {M.~D.}\ \bibnamefont {Lukin}}, \ and\ \bibinfo
  {author} {\bibfnamefont {V.}~\bibnamefont {Vuleti{\'c}}},\ }\href@noop {}
  {\bibfield  {journal} {\bibinfo  {journal} {Nature}\ }\textbf {\bibinfo
  {volume} {502}},\ \bibinfo {pages} {71} (\bibinfo {year} {2013})}\BibitemShut
  {NoStop}%
\bibitem [{\citenamefont {He}\ \emph {et~al.}(2014)\citenamefont {He},
  \citenamefont {Sharypov}, \citenamefont {Sheng}, \citenamefont {Simon},\ and\
  \citenamefont {Xiao}}]{he2014two}%
  \BibitemOpen
  \bibfield  {author} {\bibinfo {author} {\bibfnamefont {B.}~\bibnamefont
  {He}}, \bibinfo {author} {\bibfnamefont {A.V.}~\bibnamefont {Sharypov}},
  \bibinfo {author} {\bibfnamefont {J.}~\bibnamefont {Sheng}}, \bibinfo
  {author} {\bibfnamefont {C.}~\bibnamefont {Simon}}, \ and\ \bibinfo {author}
  {\bibfnamefont {M.}~\bibnamefont {Xiao}},\ }\href@noop {} {\bibfield
  {journal} {\bibinfo  {journal} {Physical review letters}\ }\textbf {\bibinfo
  {volume} {112}},\ \bibinfo {pages} {133606} (\bibinfo {year}
  {2014})}\BibitemShut {NoStop}%
\bibitem [{\citenamefont {Tiarks}\ \emph {et~al.}(2014)\citenamefont {Tiarks},
  \citenamefont {Baur}, \citenamefont {Schneider}, \citenamefont {D{\"u}rr},\
  and\ \citenamefont {Rempe}}]{tiarks2014single}%
  \BibitemOpen
  \bibfield  {author} {\bibinfo {author} {\bibfnamefont {D.}~\bibnamefont
  {Tiarks}}, \bibinfo {author} {\bibfnamefont {S.}~\bibnamefont {Baur}},
  \bibinfo {author} {\bibfnamefont {K.}~\bibnamefont {Schneider}}, \bibinfo
  {author} {\bibfnamefont {S.}~\bibnamefont {D{\"u}rr}}, \ and\ \bibinfo
  {author} {\bibfnamefont {G.}~\bibnamefont {Rempe}},\ }\href@noop {}
  {\bibfield  {journal} {\bibinfo  {journal} {Physical review letters}\
  }\textbf {\bibinfo {volume} {113}},\ \bibinfo {pages} {053602} (\bibinfo
  {year} {2014})}\BibitemShut {NoStop}%
\bibitem [{\citenamefont {Maghrebi}\ \emph {et~al.}(2015)\citenamefont
  {Maghrebi}, \citenamefont {Gullans}, \citenamefont {Bienias}, \citenamefont
  {Choi}, \citenamefont {Martin}, \citenamefont {Firstenberg}, \citenamefont
  {Lukin}, \citenamefont {B{\"u}chler},\ and\ \citenamefont
  {Gorshkov}}]{maghrebi2015coulomb}%
  \BibitemOpen
  \bibfield  {author} {\bibinfo {author} {\bibfnamefont {M.F}~\bibnamefont
  {Maghrebi}}, \bibinfo {author} {\bibfnamefont {M.J.}~\bibnamefont {Gullans}},
  \bibinfo {author} {\bibfnamefont {P.}~\bibnamefont {Bienias}}, \bibinfo
  {author} {\bibfnamefont {S.}~\bibnamefont {Choi}}, \bibinfo {author}
  {\bibfnamefont {I.}~\bibnamefont {Martin}}, \bibinfo {author} {\bibfnamefont
  {O.}~\bibnamefont {Firstenberg}}, \bibinfo {author} {\bibfnamefont {M.~D.}\
  \bibnamefont {Lukin}}, \bibinfo {author} {\bibfnamefont {H.~P.}~\bibnamefont
  {B{\"u}chler}}, \ and\ \bibinfo {author} {\bibfnamefont {A.~V.}\ \bibnamefont
  {Gorshkov}},\ }\href@noop {} {\bibfield  {journal} {\bibinfo  {journal}
  {Physical review letters}\ }\textbf {\bibinfo {volume} {115}},\ \bibinfo
  {pages} {123601} (\bibinfo {year} {2015})}\BibitemShut {NoStop}%
\bibitem [{\citenamefont {Grankin}\ \emph
  {et~al.}(2016{\natexlab{a}})\citenamefont {Grankin}, \citenamefont {Brion},
  \citenamefont {Boddeda}, \citenamefont {{\'C}uk}, \citenamefont {Usmani},
  \citenamefont {Ourjoumtsev},\ and\ \citenamefont
  {Grangier}}]{grankin2016inelastic}%
  \BibitemOpen
  \bibfield  {author} {\bibinfo {author} {\bibfnamefont {A.}~\bibnamefont
  {Grankin}}, \bibinfo {author} {\bibfnamefont {E.}~\bibnamefont {Brion}},
  \bibinfo {author} {\bibfnamefont {R.}~\bibnamefont {Boddeda}}, \bibinfo
  {author} {\bibfnamefont {S.}~\bibnamefont {{\'C}uk}}, \bibinfo {author}
  {\bibfnamefont {I.}~\bibnamefont {Usmani}}, \bibinfo {author} {\bibfnamefont
  {A.}~\bibnamefont {Ourjoumtsev}}, \ and\ \bibinfo {author} {\bibfnamefont
  {P.}~\bibnamefont {Grangier}},\ }\href@noop {} {\bibfield  {journal}
  {\bibinfo  {journal} {Physical Review Letters}\ }\textbf {\bibinfo {volume}
  {117}},\ \bibinfo {pages} {253602} (\bibinfo {year}
  {2016}{\natexlab{a}})}\BibitemShut {NoStop}%
\bibitem [{\citenamefont {Gorniaczyk}\ \emph {et~al.}(2016)\citenamefont
  {Gorniaczyk}, \citenamefont {Tresp}, \citenamefont {Bienias}, \citenamefont
  {Paris-Mandoki}, \citenamefont {Li}, \citenamefont {Mirgorodskiy},
  \citenamefont {B{\"u}chler}, \citenamefont {Lesanovsky},\ and\ \citenamefont
  {Hofferberth}}]{gorniaczyk2016enhancement}%
  \BibitemOpen
  \bibfield  {author} {\bibinfo {author} {\bibfnamefont {H.}~\bibnamefont
  {Gorniaczyk}}, \bibinfo {author} {\bibfnamefont {C.}~\bibnamefont {Tresp}},
  \bibinfo {author} {\bibfnamefont {P.}~\bibnamefont {Bienias}}, \bibinfo
  {author} {\bibfnamefont {A.}~\bibnamefont {Paris-Mandoki}}, \bibinfo {author}
  {\bibfnamefont {W.}~\bibnamefont {Li}}, \bibinfo {author} {\bibfnamefont
  {I.}~\bibnamefont {Mirgorodskiy}}, \bibinfo {author} {\bibfnamefont
  {H.}~\bibnamefont {B{\"u}chler}}, \bibinfo {author} {\bibfnamefont
  {I.}~\bibnamefont {Lesanovsky}}, \ and\ \bibinfo {author} {\bibfnamefont
  {S.}~\bibnamefont {Hofferberth}},\ }\href@noop {} {\bibfield  {journal}
  {\bibinfo  {journal} {Nature communications}\ }\textbf {\bibinfo {volume}
  {7}},\ \bibinfo {pages} {12480} (\bibinfo {year} {2016})}\BibitemShut
  {NoStop}%
\bibitem [{\citenamefont {Jachymski}\ \emph
  {et~al.}(2016{\natexlab{a}})\citenamefont {Jachymski}, \citenamefont
  {Bienias},\ and\ \citenamefont {B{\"u}chler}}]{jachymski2016three}%
  \BibitemOpen
  \bibfield  {author} {\bibinfo {author} {\bibfnamefont {K.}~\bibnamefont
  {Jachymski}}, \bibinfo {author} {\bibfnamefont {P.}~\bibnamefont {Bienias}},
  \ and\ \bibinfo {author} {\bibfnamefont {H.~P.}\ \bibnamefont
  {B{\"u}chler}},\ }\href@noop {} {\bibfield  {journal} {\bibinfo  {journal}
  {Physical Review Letters}\ }\textbf {\bibinfo {volume} {117}},\ \bibinfo
  {pages} {053601} (\bibinfo {year} {2016}{\natexlab{a}})}\BibitemShut
  {NoStop}%
\bibitem [{\citenamefont {Grankin}\ \emph
  {et~al.}(2016{\natexlab{b}})\citenamefont {Grankin}, \citenamefont {Brion},
  \citenamefont {Boddeda}, \citenamefont {\ifmmode~\acute{C}\else
  \'{C}\fi{}uk}, \citenamefont {Usmani}, \citenamefont {Ourjoumtsev},\ and\
  \citenamefont {Grangier}}]{Grangier16}%
  \BibitemOpen
  \bibfield  {author} {\bibinfo {author} {\bibfnamefont {A.}~\bibnamefont
  {Grankin}}, \bibinfo {author} {\bibfnamefont {E.}~\bibnamefont {Brion}},
  \bibinfo {author} {\bibfnamefont {R.}~\bibnamefont {Boddeda}}, \bibinfo
  {author} {\bibfnamefont {S.}~\bibnamefont {\ifmmode~\acute{C}\else
  \'{C}\fi{}uk}}, \bibinfo {author} {\bibfnamefont {I.}~\bibnamefont {Usmani}},
  \bibinfo {author} {\bibfnamefont {A.}~\bibnamefont {Ourjoumtsev}}, \ and\
  \bibinfo {author} {\bibfnamefont {P.}~\bibnamefont {Grangier}},\ }\href
  {\doibase 10.1103/PhysRevLett.117.253602} {\bibfield  {journal} {\bibinfo
  {journal} {Phys. Rev. Lett.}\ }\textbf {\bibinfo {volume} {117}},\ \bibinfo
  {pages} {253602} (\bibinfo {year} {2016}{\natexlab{b}})}\BibitemShut
  {NoStop}%
\bibitem [{\citenamefont {Winchester}\ \emph {et~al.}(2017)\citenamefont
  {Winchester}, \citenamefont {Norcia}, \citenamefont {Cline},\ and\
  \citenamefont {Thompson}}]{Thompson17}%
  \BibitemOpen
  \bibfield  {author} {\bibinfo {author} {\bibfnamefont {M.~N.}\ \bibnamefont
  {Winchester}}, \bibinfo {author} {\bibfnamefont {M.~A.}\ \bibnamefont
  {Norcia}}, \bibinfo {author} {\bibfnamefont {J.~R.~K.}\ \bibnamefont
  {Cline}}, \ and\ \bibinfo {author} {\bibfnamefont {J.~K.}\ \bibnamefont
  {Thompson}},\ }\href {\doibase 10.1103/PhysRevLett.118.263601} {\bibfield
  {journal} {\bibinfo  {journal} {Phys. Rev. Lett.}\ }\textbf {\bibinfo
  {volume} {118}},\ \bibinfo {pages} {263601} (\bibinfo {year}
  {2017})}\BibitemShut {NoStop}%
\bibitem [{\citenamefont {Ye}\ \emph {et~al.}(1999)\citenamefont {Ye},
  \citenamefont {Vernooy},\ and\ \citenamefont {Kimble}}]{Kimble99}%
  \BibitemOpen
  \bibfield  {author} {\bibinfo {author} {\bibfnamefont {J.}~\bibnamefont
  {Ye}}, \bibinfo {author} {\bibfnamefont {D.~W.}\ \bibnamefont {Vernooy}}, \
  and\ \bibinfo {author} {\bibfnamefont {H.~J.}\ \bibnamefont {Kimble}},\
  }\href {\doibase 10.1103/PhysRevLett.83.4987} {\bibfield  {journal} {\bibinfo
   {journal} {Phys. Rev. Lett.}\ }\textbf {\bibinfo {volume} {83}},\ \bibinfo
  {pages} {4987} (\bibinfo {year} {1999})}\BibitemShut {NoStop}%
\bibitem [{\citenamefont {P.~Berman}()}]{Berman1994}%
  \BibitemOpen
  \bibfield  {author} {\bibinfo {editor} {\bibfnamefont {P.}~\bibnamefont
  {Berman}},\ }\href@noop {} {\emph {\bibinfo {title} {Cavity Quantum Electrodynamics}}}\
  (\bibinfo  {publisher} {Academic Press},\ \bibinfo {year}
  {1994})\BibitemShut {NoStop}
\bibitem [{\citenamefont {Sterk}\ \emph {et~al.}(2012)\citenamefont {Sterk},
  \citenamefont {Luo}, \citenamefont {Manning}, \citenamefont {Maunz},\ and\
  \citenamefont {Monroe}}]{sterk2012photon}%
  \BibitemOpen
  \bibfield  {author} {\bibinfo {author} {\bibfnamefont {J.~D.}~\bibnamefont
  {Sterk}}, \bibinfo {author} {\bibfnamefont {L.}~\bibnamefont {Luo}}, \bibinfo
  {author} {\bibfnamefont {T.~A.}~\bibnamefont {Manning}}, \bibinfo {author}
  {\bibfnamefont {P.}~\bibnamefont {Maunz}}, \ and\ \bibinfo {author}
  {\bibfnamefont {C.}~\bibnamefont {Monroe}},\ }\href@noop {} {\bibfield
  {journal} {\bibinfo  {journal} {Phys. Rev. A}\ }\textbf {\bibinfo {volume}
  {85}},\ \bibinfo {pages} {062308} (\bibinfo {year} {2012})}\BibitemShut
  {NoStop}%
\bibitem [{\citenamefont {Casabone}\ \emph {et~al.}(2015)\citenamefont
  {Casabone}, \citenamefont {Friebe}, \citenamefont {Brandstatter},
  \citenamefont {Schuppert}, \citenamefont {Blatt},\ and\ \citenamefont
  {Northup}}]{casabone2015enhanced}%
  \BibitemOpen
  \bibfield  {author} {\bibinfo {author} {\bibfnamefont {B.}~\bibnamefont
  {Casabone}}, \bibinfo {author} {\bibfnamefont {K.}~\bibnamefont {Friebe}},
  \bibinfo {author} {\bibfnamefont {B.}~\bibnamefont {Brandstatter}}, \bibinfo
  {author} {\bibfnamefont {K.}~\bibnamefont {Schuppert}}, \bibinfo {author}
  {\bibfnamefont {R.}~\bibnamefont {Blatt}}, \ and\ \bibinfo {author}
  {\bibfnamefont {T.~E.}~\bibnamefont {Northup}},\ }\href@noop {} {\bibfield
  {journal} {\bibinfo  {journal} {Phys. Rev. Lett.}\ }\textbf {\bibinfo
  {volume} {114}},\ \bibinfo {pages} {023602} (\bibinfo {year}
  {2015})}\BibitemShut {NoStop}%
\bibitem [{\citenamefont {Tischler}\ \emph {et~al.}(2005)\citenamefont
  {Tischler}, \citenamefont {Bradley}, \citenamefont {Bulovic}, \citenamefont
  {Song},\ and\ \citenamefont {Nurmikko}}]{tischler2005strong}%
  \BibitemOpen
  \bibfield  {author} {\bibinfo {author} {\bibfnamefont {J.~R.}\ \bibnamefont
  {Tischler}}, \bibinfo {author} {\bibfnamefont {M.~S.}\ \bibnamefont
  {Bradley}}, \bibinfo {author} {\bibfnamefont {V.}~\bibnamefont {Bulovic}},
  \bibinfo {author} {\bibfnamefont {J.~H.}\ \bibnamefont {Song}}, \ and\
  \bibinfo {author} {\bibfnamefont {A.}~\bibnamefont {Nurmikko}},\ }\href@noop
  {} {\bibfield  {journal} {\bibinfo  {journal} {Phys. Rev. Lett.}\ }\textbf
  {\bibinfo {volume} {95}},\ \bibinfo {pages} {036401} (\bibinfo {year}
  {2005})}\BibitemShut {NoStop}%
\bibitem [{\citenamefont {Reithmaier}\ \emph {et~al.}(2004)\citenamefont
  {Reithmaier}, \citenamefont {Sek}, \citenamefont {Loffler}, \citenamefont
  {Hofmann}, \citenamefont {Kuhn}, \citenamefont {Reitzenstein}, \citenamefont
  {Keldysh}, \citenamefont {Kulakovskii}, \citenamefont {Reinecke},\ and\
  \citenamefont {Forchel}}]{reithmaier2004strong}%
  \BibitemOpen
  \bibfield  {author} {\bibinfo {author} {\bibfnamefont {J.}~\bibnamefont
  {Reithmaier}}, \bibinfo {author} {\bibfnamefont {G.}~\bibnamefont {Sek}},
  \bibinfo {author} {\bibfnamefont {A.}~\bibnamefont {Loffler}}, \bibinfo
  {author} {\bibfnamefont {C.}~\bibnamefont {Hofmann}}, \bibinfo {author}
  {\bibfnamefont {S.}~\bibnamefont {Kuhn}}, \bibinfo {author} {\bibfnamefont
  {S.}~\bibnamefont {Reitzenstein}}, \bibinfo {author} {\bibfnamefont
  {L.}~\bibnamefont {Keldysh}}, \bibinfo {author} {\bibfnamefont
  {V.}~\bibnamefont {Kulakovskii}}, \bibinfo {author} {\bibfnamefont
  {T.}~\bibnamefont {Reinecke}}, \ and\ \bibinfo {author} {\bibfnamefont
  {A.}~\bibnamefont {Forchel}},\ }\href@noop {} {\bibfield  {journal} {\bibinfo
   {journal} {Nature}\ }\textbf {\bibinfo {volume} {432}},\ \bibinfo {pages}
  {197} (\bibinfo {year} {2004})}\BibitemShut {NoStop}%
\bibitem [{\citenamefont {Park}\ \emph {et~al.}(2006)\citenamefont {Park},
  \citenamefont {Cook},\ and\ \citenamefont {Wang}}]{park2006cavity}%
  \BibitemOpen
  \bibfield  {author} {\bibinfo {author} {\bibfnamefont {Y.-S.}\ \bibnamefont
  {Park}}, \bibinfo {author} {\bibfnamefont {A.~K.}\ \bibnamefont {Cook}}, \
  and\ \bibinfo {author} {\bibfnamefont {H.}~\bibnamefont {Wang}},\ }\href@noop
  {} {\bibfield  {journal} {\bibinfo  {journal} {Nano Lett.}\ }\textbf
  {\bibinfo {volume} {6}},\ \bibinfo {pages} {2075} (\bibinfo {year}
  {2006})}\BibitemShut {NoStop}%
\bibitem [{\citenamefont {Haroche}\ and\ \citenamefont
  {Raimond}(1985)}]{haroche1985radiative}%
  \BibitemOpen
  \bibfield  {author} {\bibinfo {author} {\bibfnamefont {S.}~\bibnamefont
  {Haroche}}\ and\ \bibinfo {author} {\bibfnamefont {J.}~\bibnamefont
  {Raimond}},\ }\href@noop {} {\bibfield  {journal} {\bibinfo  {journal} {Adv.
  At. Mol. Phys.}\ }\textbf {\bibinfo {volume} {20}},\ \bibinfo {pages} {347}
  (\bibinfo {year} {1985})}\BibitemShut {NoStop}%
\bibitem [{\citenamefont {Thompson}\ \emph {et~al.}(1992)\citenamefont
  {Thompson}, \citenamefont {Rempe},\ and\ \citenamefont
  {Kimble}}]{thompson1992observation}%
  \BibitemOpen
  \bibfield  {author} {\bibinfo {author} {\bibfnamefont {R.~J.}~\bibnamefont
  {Thompson}}, \bibinfo {author} {\bibfnamefont {G.}~\bibnamefont {Rempe}}, \
  and\ \bibinfo {author} {\bibfnamefont {H.~J.}~\bibnamefont {Kimble}},\
  }\href@noop {} {\bibfield  {journal} {\bibinfo  {journal} {Phys. Rev. Lett.}\
  }\textbf {\bibinfo {volume} {68}},\ \bibinfo {pages} {1132} (\bibinfo {year}
  {1992})}\BibitemShut {NoStop}%
\bibitem [{\citenamefont {Aoki}\ \emph {et~al.}(2006)\citenamefont {Aoki},
  \citenamefont {Dayan}, \citenamefont {Wilcut}, \citenamefont {Bowen},
  \citenamefont {Parkins}, \citenamefont {Kippenberg}, \citenamefont {Vahala},\
  and\ \citenamefont {Kimble}}]{aoki2006observation}%
  \BibitemOpen
  \bibfield  {author} {\bibinfo {author} {\bibfnamefont {T.}~\bibnamefont
  {Aoki}}, \bibinfo {author} {\bibfnamefont {B.}~\bibnamefont {Dayan}},
  \bibinfo {author} {\bibfnamefont {E.}~\bibnamefont {Wilcut}}, \bibinfo
  {author} {\bibfnamefont {W.~P.}\ \bibnamefont {Bowen}}, \bibinfo {author}
  {\bibfnamefont {A.~S.}\ \bibnamefont {Parkins}}, \bibinfo {author}
  {\bibfnamefont {T.}~\bibnamefont {Kippenberg}}, \bibinfo {author}
  {\bibfnamefont {K.}~\bibnamefont {Vahala}}, \ and\ \bibinfo {author}
  {\bibfnamefont {H.}~\bibnamefont {Kimble}},\ }\href@noop {} {\bibfield
  {journal} {\bibinfo  {journal} {Nature}\ }\textbf {\bibinfo {volume} {443}},\
  \bibinfo {pages} {671} (\bibinfo {year} {2006})}\BibitemShut {NoStop}%
\bibitem [{\citenamefont {Yoshie}\ \emph {et~al.}(2004)\citenamefont {Yoshie},
  \citenamefont {Scherer}, \citenamefont {Hendrickson}, \citenamefont
  {Khitrova}, \citenamefont {Gibbs}, \citenamefont {Rupper}, \citenamefont
  {Ell}, \citenamefont {Shchekin},\ and\ \citenamefont
  {Deppe}}]{yoshie2004vacuum}%
  \BibitemOpen
  \bibfield  {author} {\bibinfo {author} {\bibfnamefont {T.}~\bibnamefont
  {Yoshie}}, \bibinfo {author} {\bibfnamefont {A.}~\bibnamefont {Scherer}},
  \bibinfo {author} {\bibfnamefont {J.}~\bibnamefont {Hendrickson}}, \bibinfo
  {author} {\bibfnamefont {G.}~\bibnamefont {Khitrova}}, \bibinfo {author}
  {\bibfnamefont {H.}~\bibnamefont {Gibbs}}, \bibinfo {author} {\bibfnamefont
  {G.}~\bibnamefont {Rupper}}, \bibinfo {author} {\bibfnamefont
  {C.}~\bibnamefont {Ell}}, \bibinfo {author} {\bibfnamefont {O.}~\bibnamefont
  {Shchekin}}, \ and\ \bibinfo {author} {\bibfnamefont {D.}~\bibnamefont
  {Deppe}},\ }\href@noop {} {\bibfield  {journal} {\bibinfo  {journal}
  {Nature}\ }\textbf {\bibinfo {volume} {432}},\ \bibinfo {pages} {200}
  (\bibinfo {year} {2004})}\BibitemShut {NoStop}%
\bibitem [{\citenamefont {Haas}\ \emph {et~al.}(2014)\citenamefont {Haas},
  \citenamefont {Volz}, \citenamefont {Gehr}, \citenamefont {Reichel},\ and\
  \citenamefont {Esteve}}]{haas2014entangled}%
  \BibitemOpen
  \bibfield  {author} {\bibinfo {author} {\bibfnamefont {F.}~\bibnamefont
  {Haas}}, \bibinfo {author} {\bibfnamefont {J.}~\bibnamefont {Volz}}, \bibinfo
  {author} {\bibfnamefont {R.}~\bibnamefont {Gehr}}, \bibinfo {author}
  {\bibfnamefont {J.}~\bibnamefont {Reichel}}, \ and\ \bibinfo {author}
  {\bibfnamefont {J.}~\bibnamefont {Esteve}},\ }\href@noop {} {\bibfield
  {journal} {\bibinfo  {journal} {Science}\ }\textbf {\bibinfo {volume}
  {344}},\ \bibinfo {pages} {180} (\bibinfo {year} {2014})}\BibitemShut
  {NoStop}%
\bibitem [{\citenamefont {Blais}\ \emph {et~al.}(2004)\citenamefont {Blais},
  \citenamefont {Huang}, \citenamefont {Wallraff}, \citenamefont {Girvin},\
  and\ \citenamefont {Schoelkopf}}]{blais2004cavity}%
  \BibitemOpen
  \bibfield  {author} {\bibinfo {author} {\bibfnamefont {A.}~\bibnamefont
  {Blais}}, \bibinfo {author} {\bibfnamefont {R.~S.}\ \bibnamefont {Huang}},
  \bibinfo {author} {\bibfnamefont {A.}~\bibnamefont {Wallraff}}, \bibinfo
  {author} {\bibfnamefont {S.~M.}~\bibnamefont {Girvin}}, \ and\ \bibinfo {author}
  {\bibfnamefont {R.~J.}\ \bibnamefont {Schoelkopf}},\ }\href@noop {}
  {\bibfield  {journal} {\bibinfo  {journal} {Phys. Rev. A}\ }\textbf {\bibinfo
  {volume} {69}},\ \bibinfo {pages} {062320} (\bibinfo {year}
  {2004})}\BibitemShut {NoStop}%
\bibitem [{\citenamefont {Sheng}\ \emph {et~al.}(2016)\citenamefont {Sheng},
  \citenamefont {Chao},\ and\ \citenamefont {Shaffer}}]{Sheng2016}%
  \BibitemOpen
  \bibfield  {author} {\bibinfo {author} {\bibfnamefont {J.}~\bibnamefont
  {Sheng}}, \bibinfo {author} {\bibfnamefont {Y.}~\bibnamefont {Chao}}, \ and\
  \bibinfo {author} {\bibfnamefont {J.~P.}\ \bibnamefont {Shaffer}},\ }\href
  {\doibase 10.1103/PhysRevLett.117.103201} {\bibfield  {journal} {\bibinfo
  {journal} {Phys. Rev. Lett.}\ }\textbf {\bibinfo {volume} {117}},\ \bibinfo
  {pages} {103201} (\bibinfo {year} {2016})}\BibitemShut {NoStop}%
\bibitem [{\citenamefont {Fleischhauer}\ \emph {et~al.}(2005)\citenamefont
  {Fleischhauer}, \citenamefont {Imamoglu},\ and\ \citenamefont
  {Marangos}}]{fleischhauer2005electromagnetically}%
  \BibitemOpen
  \bibfield  {author} {\bibinfo {author} {\bibfnamefont {M.}~\bibnamefont
  {Fleischhauer}}, \bibinfo {author} {\bibfnamefont {A.}~\bibnamefont
  {Imamoglu}}, \ and\ \bibinfo {author} {\bibfnamefont {J.~P.}\ \bibnamefont
  {Marangos}},\ }\href@noop {} {\bibfield  {journal} {\bibinfo  {journal} {Rev.
  Mod. Phys.}\ }\textbf {\bibinfo {volume} {77}},\ \bibinfo {pages} {633}
  (\bibinfo {year} {2005})}\BibitemShut {NoStop}%
\bibitem [{\citenamefont {Mohapatra}\ \emph {et~al.}(2007)\citenamefont
  {Mohapatra}, \citenamefont {Jackson},\ and\ \citenamefont
  {Adams}}]{mohapatra2007coherent}%
  \BibitemOpen
  \bibfield  {author} {\bibinfo {author} {\bibfnamefont {A.~K.}~\bibnamefont
  {Mohapatra}}, \bibinfo {author} {\bibfnamefont {T.~R.}~\bibnamefont {Jackson}},
  \ and\ \bibinfo {author} {\bibfnamefont {C.~S.}~\bibnamefont {Adams}},\
  }\href@noop {} {\bibfield  {journal} {\bibinfo  {journal} {Phys. Rev. Lett.}\
  }\textbf {\bibinfo {volume} {98}},\ \bibinfo {pages} {113003} (\bibinfo
  {year} {2007})}\BibitemShut {NoStop}%
\bibitem [{\citenamefont {Parigi}\ \emph {et~al.}(2012)\citenamefont {Parigi},
  \citenamefont {Bimbard}, \citenamefont {Stanojevic}, \citenamefont
  {Hilliard}, \citenamefont {Nogrette}, \citenamefont {Tualle-Brouri},
  \citenamefont {Ourjoumtsev},\ and\ \citenamefont
  {Grangier}}]{parigi2012observation}%
  \BibitemOpen
  \bibfield  {author} {\bibinfo {author} {\bibfnamefont {V.}~\bibnamefont
  {Parigi}}, \bibinfo {author} {\bibfnamefont {E.}~\bibnamefont {Bimbard}},
  \bibinfo {author} {\bibfnamefont {J.}~\bibnamefont {Stanojevic}}, \bibinfo
  {author} {\bibfnamefont {A.~J.}\ \bibnamefont {Hilliard}}, \bibinfo {author}
  {\bibfnamefont {F.}~\bibnamefont {Nogrette}}, \bibinfo {author}
  {\bibfnamefont {R.}~\bibnamefont {Tualle-Brouri}}, \bibinfo {author}
  {\bibfnamefont {A.}~\bibnamefont {Ourjoumtsev}}, \ and\ \bibinfo {author}
  {\bibfnamefont {P.}~\bibnamefont {Grangier}},\ }\href@noop {} {\bibfield
  {journal} {\bibinfo  {journal} {Phys. Rev. Lett.}\ }\textbf {\bibinfo
  {volume} {109}},\ \bibinfo {pages} {233602} (\bibinfo {year}
  {2012})}\BibitemShut {NoStop}%
\bibitem [{\citenamefont {Boddeda}\ \emph {et~al.}(2016)\citenamefont
  {Boddeda}, \citenamefont {Usmani}, \citenamefont {Bimbard}, \citenamefont
  {Grankin}, \citenamefont {Ourjoumtsev}, \citenamefont {Brion},\ and\
  \citenamefont {Grangier}}]{boddeda2016rydberg}%
  \BibitemOpen
  \bibfield  {author} {\bibinfo {author} {\bibfnamefont {R.}~\bibnamefont
  {Boddeda}}, \bibinfo {author} {\bibfnamefont {I.}~\bibnamefont {Usmani}},
  \bibinfo {author} {\bibfnamefont {E.}~\bibnamefont {Bimbard}}, \bibinfo
  {author} {\bibfnamefont {A.}~\bibnamefont {Grankin}}, \bibinfo {author}
  {\bibfnamefont {A.}~\bibnamefont {Ourjoumtsev}}, \bibinfo {author}
  {\bibfnamefont {E.}~\bibnamefont {Brion}}, \ and\ \bibinfo {author}
  {\bibfnamefont {P.}~\bibnamefont {Grangier}},\ }\href@noop {} {\bibfield
  {journal} {\bibinfo  {journal} {J. Phys.B}\ }\textbf {\bibinfo {volume}
  {49}},\ \bibinfo {pages} {084005} (\bibinfo {year} {2016})}\BibitemShut
  {NoStop}%
\bibitem [{\citenamefont {Ningyuan}\ \emph {et~al.}(2016)\citenamefont
  {Ningyuan}, \citenamefont {Georgakopoulos}, \citenamefont {Ryou},
  \citenamefont {Schine}, \citenamefont {Sommer},\ and\ \citenamefont
  {Simon}}]{ningyuan2016observation}%
  \BibitemOpen
  \bibfield  {author} {\bibinfo {author} {\bibfnamefont {J.}~\bibnamefont
  {Ningyuan}}, \bibinfo {author} {\bibfnamefont {A.}~\bibnamefont
  {Georgakopoulos}}, \bibinfo {author} {\bibfnamefont {A.}~\bibnamefont
  {Ryou}}, \bibinfo {author} {\bibfnamefont {N.}~\bibnamefont {Schine}},
  \bibinfo {author} {\bibfnamefont {A.}~\bibnamefont {Sommer}}, \ and\ \bibinfo
  {author} {\bibfnamefont {J.}~\bibnamefont {Simon}},\ }\href@noop {}
  {\bibfield  {journal} {\bibinfo  {journal} {Phys. Rev. A}\ }\textbf {\bibinfo
  {volume} {93}},\ \bibinfo {pages} {041802} (\bibinfo {year}
  {2016})}\BibitemShut {NoStop}%
\bibitem [{\citenamefont {Hernandez}\ \emph {et~al.}(2007)\citenamefont
  {Hernandez}, \citenamefont {Zhang},\ and\ \citenamefont {Zhu}}]{2007vacuum}%
  \BibitemOpen
  \bibfield  {author} {\bibinfo {author} {\bibfnamefont {G.}~\bibnamefont
  {Hernandez}}, \bibinfo {author} {\bibfnamefont {J.}~\bibnamefont {Zhang}}, \
  and\ \bibinfo {author} {\bibfnamefont {Y.}~\bibnamefont {Zhu}},\ }\href@noop
  {} {\bibfield  {journal} {\bibinfo  {journal} {Phys. Rev. A}\ }\textbf
  {\bibinfo {volume} {76}},\ \bibinfo {pages} {053814} (\bibinfo {year}
  {2007})}\BibitemShut {NoStop}%
\bibitem [{\citenamefont {Wu}\ \emph {et~al.}(2008)\citenamefont {Wu},
  \citenamefont {Gea-Banacloche},\ and\ \citenamefont
  {Xiao}}]{wu2008observation}%
  \BibitemOpen
  \bibfield  {author} {\bibinfo {author} {\bibfnamefont {H.}~\bibnamefont
  {Wu}}, \bibinfo {author} {\bibfnamefont {J.}~\bibnamefont {Gea-Banacloche}},
  \ and\ \bibinfo {author} {\bibfnamefont {M.}~\bibnamefont {Xiao}},\
  }\href@noop {} {\bibfield  {journal} {\bibinfo  {journal} {Phys. Rev. Lett.}\
  }\textbf {\bibinfo {volume} {100}},\ \bibinfo {pages} {173602} (\bibinfo
  {year} {2008})}\BibitemShut {NoStop}%
\bibitem [{\citenamefont {Mucke}\ \emph {et~al.}(2010)\citenamefont {Mucke},
  \citenamefont {Figueroa}, \citenamefont {Bochmann}, \citenamefont {Hahn},
  \citenamefont {Murr}, \citenamefont {Ritter}, \citenamefont {Villas-Boas},\
  and\ \citenamefont {Rempe}}]{mucke2010electromagnetically}%
  \BibitemOpen
  \bibfield  {author} {\bibinfo {author} {\bibfnamefont {M.}~\bibnamefont
  {Mucke}}, \bibinfo {author} {\bibfnamefont {E.}~\bibnamefont {Figueroa}},
  \bibinfo {author} {\bibfnamefont {J.}~\bibnamefont {Bochmann}}, \bibinfo
  {author} {\bibfnamefont {C.}~\bibnamefont {Hahn}}, \bibinfo {author}
  {\bibfnamefont {K.}~\bibnamefont {Murr}}, \bibinfo {author} {\bibfnamefont
  {S.}~\bibnamefont {Ritter}}, \bibinfo {author} {\bibfnamefont {C.~J.}\
  \bibnamefont {Villas-Boas}}, \ and\ \bibinfo {author} {\bibfnamefont
  {G.}~\bibnamefont {Rempe}},\ }\href@noop {} {\bibfield  {journal} {\bibinfo
  {journal} {Nature}\ }\textbf {\bibinfo {volume} {465}},\ \bibinfo {pages}
  {755} (\bibinfo {year} {2010})}\BibitemShut {NoStop}%
\bibitem [{\citenamefont {Albert}\ \emph {et~al.}(2011)\citenamefont {Albert},
  \citenamefont {Dantan},\ and\ \citenamefont {Drewsen}}]{albert2011cavity}%
  \BibitemOpen
  \bibfield  {author} {\bibinfo {author} {\bibfnamefont {M.}~\bibnamefont
  {Albert}}, \bibinfo {author} {\bibfnamefont {A.}~\bibnamefont {Dantan}}, \
  and\ \bibinfo {author} {\bibfnamefont {M.}~\bibnamefont {Drewsen}},\
  }\href@noop {} {\bibfield  {journal} {\bibinfo  {journal} {Nature Photonics}\
  }\textbf {\bibinfo {volume} {5}},\ \bibinfo {pages} {633} (\bibinfo {year}
  {2011})}\BibitemShut {NoStop}%
\bibitem [{\citenamefont {Jia}\ \emph {et~al.}()\citenamefont {Jia},
  \citenamefont {Schine}, \citenamefont {Georgakopoulos}, \citenamefont {Ryou},
  \citenamefont {Sommer},\ and\ \citenamefont {Simon}}]{Simon2017}%
  \BibitemOpen
  \bibfield  {author} {\bibinfo {author} {\bibfnamefont {N.}~\bibnamefont
  {Jia}}, \bibinfo {author} {\bibfnamefont {N.}~\bibnamefont {Schine}},
  \bibinfo {author} {\bibfnamefont {A.}~\bibnamefont {Georgakopoulos}},
  \bibinfo {author} {\bibfnamefont {A.}~\bibnamefont {Ryou}}, \bibinfo {author}
  {\bibfnamefont {A.}~\bibnamefont {Sommer}}, \ and\ \bibinfo {author}
  {\bibfnamefont {J.}~\bibnamefont {Simon}},\ }\href@noop {} {\ }\Eprint
  {http://arxiv.org/abs/1705.07475} {arXiv:1705.07475 [cond-mat.quant-gas]}
  \BibitemShut {NoStop}%
\bibitem [{\citenamefont {Bienias}\ \emph {et~al.}(2014)\citenamefont
  {Bienias}, \citenamefont {Choi}, \citenamefont {Firstenberg}, \citenamefont
  {Maghrebi}, \citenamefont {Gullans}, \citenamefont {Lukin}, \citenamefont
  {Gorshkov},\ and\ \citenamefont {B\"uchler}}]{Buchler14}%
  \BibitemOpen
  \bibfield  {author} {\bibinfo {author} {\bibfnamefont {P.}~\bibnamefont
  {Bienias}}, \bibinfo {author} {\bibfnamefont {S.}~\bibnamefont {Choi}},
  \bibinfo {author} {\bibfnamefont {O.}~\bibnamefont {Firstenberg}}, \bibinfo
  {author} {\bibfnamefont {M.~F.}\ \bibnamefont {Maghrebi}}, \bibinfo {author}
  {\bibfnamefont {M.}~\bibnamefont {Gullans}}, \bibinfo {author} {\bibfnamefont
  {M.~D.}\ \bibnamefont {Lukin}}, \bibinfo {author} {\bibfnamefont {A.~V.}\
  \bibnamefont {Gorshkov}}, \ and\ \bibinfo {author} {\bibfnamefont {H.~P.}\
  \bibnamefont {B\"uchler}},\ }\href {\doibase 10.1103/PhysRevA.90.053804}
  {\bibfield  {journal} {\bibinfo  {journal} {Phys. Rev. A}\ }\textbf {\bibinfo
  {volume} {90}},\ \bibinfo {pages} {053804} (\bibinfo {year}
  {2014})}\BibitemShut {NoStop}%
\bibitem [{\citenamefont {Jachymski}\ \emph
  {et~al.}(2016{\natexlab{b}})\citenamefont {Jachymski}, \citenamefont
  {Bienias},\ and\ \citenamefont {B\"uchler}}]{Buchler16}%
  \BibitemOpen
  \bibfield  {author} {\bibinfo {author} {\bibfnamefont {K.}~\bibnamefont
  {Jachymski}}, \bibinfo {author} {\bibfnamefont {P.}~\bibnamefont {Bienias}},
  \ and\ \bibinfo {author} {\bibfnamefont {H.~P.}\ \bibnamefont {B\"uchler}},\
  }\href {\doibase 10.1103/PhysRevLett.117.053601} {\bibfield  {journal}
  {\bibinfo  {journal} {Phys. Rev. Lett.}\ }\textbf {\bibinfo {volume} {117}},\
  \bibinfo {pages} {053601} (\bibinfo {year} {2016}{\natexlab{b}})}\BibitemShut
  {NoStop}%
\bibitem [{\citenamefont {Kumar}\ \emph {et~al.}(2016)\citenamefont {Kumar},
  \citenamefont {Sheng}, \citenamefont {Sedlacek}, \citenamefont {Fan},\ and\
  \citenamefont {Shaffer}}]{kumar2016collective}%
  \BibitemOpen
  \bibfield  {author} {\bibinfo {author} {\bibfnamefont {S.}~\bibnamefont
  {Kumar}}, \bibinfo {author} {\bibfnamefont {J.}~\bibnamefont {Sheng}},
  \bibinfo {author} {\bibfnamefont {J.~A.}\ \bibnamefont {Sedlacek}}, \bibinfo
  {author} {\bibfnamefont {H.}~\bibnamefont {Fan}}, \ and\ \bibinfo {author}
  {\bibfnamefont {J.~P.}\ \bibnamefont {Shaffer}},\ }\href@noop {} {\bibfield
  {journal} {\bibinfo  {journal} {J. Phys.B}\ }\textbf {\bibinfo {volume}
  {49}},\ \bibinfo {pages} {064014} (\bibinfo {year} {2016})}\BibitemShut
  {NoStop}%
\bibitem [{\citenamefont {Sedlacek}\ \emph {et~al.}(2016)\citenamefont
  {Sedlacek}, \citenamefont {Kim}, \citenamefont {Rittenhouse}, \citenamefont
  {Weck}, \citenamefont {Sadeghpour},\ and\ \citenamefont
  {Shaffer}}]{sedlacek2016electric}%
  \BibitemOpen
  \bibfield  {author} {\bibinfo {author} {\bibfnamefont {J.~A.}~\bibnamefont
  {Sedlacek}}, \bibinfo {author} {\bibfnamefont {E.}~\bibnamefont {Kim}},
  \bibinfo {author} {\bibfnamefont {S.~T.}~\bibnamefont {Rittenhouse}}, \bibinfo
  {author} {\bibfnamefont {P.~F.}~\bibnamefont {Weck}}, \bibinfo {author}
  {\bibfnamefont {H.~R.}~\bibnamefont {Sadeghpour}}, \ and\ \bibinfo {author}
  {\bibfnamefont {J.~P.}~\bibnamefont {Shaffer}},\ }\href@noop {} {\bibfield
  {journal} {\bibinfo  {journal} {Phys. Rev. Lett.}\ }\textbf {\bibinfo
  {volume} {116}},\ \bibinfo {pages} {133201} (\bibinfo {year}
  {2016})}\BibitemShut {NoStop}%
\bibitem [{\citenamefont {Walls}\ and\ \citenamefont
  {Milburn}(2007)}]{walls2007quantum}%
  \BibitemOpen
  \bibfield  {author} {\bibinfo {author} {\bibfnamefont {D.~F.}\ \bibnamefont
  {Walls}}\ and\ \bibinfo {author} {\bibfnamefont {G.~J.}\ \bibnamefont
  {Milburn}},\ }\href@noop {} {\emph {\bibinfo {title} {Quantum optics}}}\
  (\bibinfo  {publisher} {Springer Science \& Business Media},\ \bibinfo {year}
  {2007})\BibitemShut {NoStop}%
\bibitem [{\citenamefont {Gea-Banacloche}\ \emph {et~al.}(1995)\citenamefont
  {Gea-Banacloche}, \citenamefont {Li}, \citenamefont {Jin},\ and\
  \citenamefont {Xiao}}]{Min1995}%
  \BibitemOpen
  \bibfield  {author} {\bibinfo {author} {\bibfnamefont {J.}~\bibnamefont
  {Gea-Banacloche}}, \bibinfo {author} {\bibfnamefont {Y.~Q.}\ \bibnamefont
  {Li}}, \bibinfo {author} {\bibfnamefont {S.~Z.}\ \bibnamefont {Jin}}, \ and\
  \bibinfo {author} {\bibfnamefont {M.}~\bibnamefont {Xiao}},\ }\href {\doibase
  10.1103/PhysRevA.51.576} {\bibfield  {journal} {\bibinfo  {journal} {Phys.
  Rev. A}\ }\textbf {\bibinfo {volume} {51}},\ \bibinfo {pages} {576} (\bibinfo
  {year} {1995})}\BibitemShut {NoStop}%
\bibitem [{\citenamefont {Sheng}\ \emph {et~al.}(2011)\citenamefont {Sheng},
  \citenamefont {Wu}, \citenamefont {Mumba}, \citenamefont {Gea-Banacloche},\
  and\ \citenamefont {Xiao}}]{sheng2011understanding}%
  \BibitemOpen
  \bibfield  {author} {\bibinfo {author} {\bibfnamefont {J.}~\bibnamefont
  {Sheng}}, \bibinfo {author} {\bibfnamefont {H.}~\bibnamefont {Wu}}, \bibinfo
  {author} {\bibfnamefont {M.}~\bibnamefont {Mumba}}, \bibinfo {author}
  {\bibfnamefont {J.}~\bibnamefont {Gea-Banacloche}}, \ and\ \bibinfo {author}
  {\bibfnamefont {M.}~\bibnamefont {Xiao}},\ }\href@noop {} {\bibfield
  {journal} {\bibinfo  {journal} {Phys. Rev. A}\ }\textbf {\bibinfo {volume}
  {83}},\ \bibinfo {pages} {023829} (\bibinfo {year} {2011})}\BibitemShut
  {NoStop}%
\bibitem [{\citenamefont {Léonard}\ \emph {et~al.}(2014)\citenamefont
  {Léonard}, \citenamefont {Lee}, \citenamefont {Morales}, \citenamefont
  {Karg}, \citenamefont {Esslinger},\ and\ \citenamefont
  {Donner}}]{Esslinger14}%
  \BibitemOpen
  \bibfield  {author} {\bibinfo {author} {\bibfnamefont {J.}~\bibnamefont
  {Léonard}}, \bibinfo {author} {\bibfnamefont {M.}~\bibnamefont {Lee}},
  \bibinfo {author} {\bibfnamefont {A.}~\bibnamefont {Morales}}, \bibinfo
  {author} {\bibfnamefont {T.~M.}\ \bibnamefont {Karg}}, \bibinfo {author}
  {\bibfnamefont {T.}~\bibnamefont {Esslinger}}, \ and\ \bibinfo {author}
  {\bibfnamefont {T.}~\bibnamefont {Donner}},\ }\href
  {http://stacks.iop.org/1367-2630/16/i=9/a=093028} {\bibfield  {journal}
  {\bibinfo  {journal} {New Journal of Physics}\ }\textbf {\bibinfo {volume}
  {16}},\ \bibinfo {pages} {093028} (\bibinfo {year} {2014})}\BibitemShut
  {NoStop}%
\bibitem [{\citenamefont {Milani}\ \emph {et~al.}(2017)\citenamefont {Milani},
  \citenamefont {Rauf}, \citenamefont {Barbieri}, \citenamefont {Bregolin},
  \citenamefont {Pizzocaro}, \citenamefont {Thoumany}, \citenamefont {Levi},\
  and\ \citenamefont {Calonico}}]{Milani17}%
  \BibitemOpen
  \bibfield  {author} {\bibinfo {author} {\bibfnamefont {G.}~\bibnamefont
  {Milani}}, \bibinfo {author} {\bibfnamefont {B.}~\bibnamefont {Rauf}},
  \bibinfo {author} {\bibfnamefont {P.}~\bibnamefont {Barbieri}}, \bibinfo
  {author} {\bibfnamefont {F.}~\bibnamefont {Bregolin}}, \bibinfo {author}
  {\bibfnamefont {M.}~\bibnamefont {Pizzocaro}}, \bibinfo {author}
  {\bibfnamefont {P.}~\bibnamefont {Thoumany}}, \bibinfo {author}
  {\bibfnamefont {F.}~\bibnamefont {Levi}}, \ and\ \bibinfo {author}
  {\bibfnamefont {D.}~\bibnamefont {Calonico}},\ }\href {\doibase
  10.1364/OL.42.001970} {\bibfield  {journal} {\bibinfo  {journal} {Opt.
  Lett.}\ }\textbf {\bibinfo {volume} {42}},\ \bibinfo {pages} {1970} (\bibinfo
  {year} {2017})}\BibitemShut {NoStop}%
\bibitem [{\citenamefont {Hui}(2014)}]{hui2014cavity}%
  \BibitemOpen
  \bibfield  {author} {\bibinfo {author} {\bibfnamefont {L.~C.}\ \bibnamefont
  {Hui}},\ }\emph {\bibinfo {title} {Cavity QED with Rydberg Excitations}},\
  \href@noop {} {Ph.D. thesis} (\bibinfo {year} {2014})\BibitemShut {NoStop}%
\bibitem [{\citenamefont {Gripp}\ \emph {et~al.}(1997)\citenamefont {Gripp},
  \citenamefont {Mielke},\ and\ \citenamefont {Orozco}}]{gripp1997evolution}%
  \BibitemOpen
  \bibfield  {author} {\bibinfo {author} {\bibfnamefont {J.}~\bibnamefont
  {Gripp}}, \bibinfo {author} {\bibfnamefont {S.~L.}~\bibnamefont {Mielke}}, \
  and\ \bibinfo {author} {\bibfnamefont {L.~A.}~\bibnamefont {Orozco}},\
  }\href@noop {} {\bibfield  {journal} {\bibinfo  {journal} {Phys. Rev. A}\
  }\textbf {\bibinfo {volume} {56}},\ \bibinfo {pages} {3262} (\bibinfo {year}
  {1997})}\BibitemShut {NoStop}%
\bibitem [{\citenamefont {Lukin}\ \emph {et~al.}(1998)\citenamefont {Lukin},
  \citenamefont {Fleischhauer}, \citenamefont {Scully},\ and\ \citenamefont
  {Velichansky}}]{lukin1998intracavity}%
  \BibitemOpen
  \bibfield  {author} {\bibinfo {author} {\bibfnamefont {M.~D.}\ \bibnamefont
  {Lukin}}, \bibinfo {author} {\bibfnamefont {M.}~\bibnamefont {Fleischhauer}},
  \bibinfo {author} {\bibfnamefont {M.~O.}\ \bibnamefont {Scully}}, \ and\
  \bibinfo {author} {\bibfnamefont {V.~L.}\ \bibnamefont {Velichansky}},\
  }\href@noop {} {\bibfield  {journal} {\bibinfo  {journal} {Opt. Lett.}\
  }\textbf {\bibinfo {volume} {23}},\ \bibinfo {pages} {295} (\bibinfo {year}
  {1998})}\BibitemShut {NoStop}%
\bibitem [{\citenamefont {Wang}\ \emph {et~al.}(2000)\citenamefont {Wang},
  \citenamefont {Goorskey}, \citenamefont {Burkett},\ and\ \citenamefont
  {Xiao}}]{wang2000cavity}%
  \BibitemOpen
  \bibfield  {author} {\bibinfo {author} {\bibfnamefont {H.}~\bibnamefont
  {Wang}}, \bibinfo {author} {\bibfnamefont {D.}~\bibnamefont {Goorskey}},
  \bibinfo {author} {\bibfnamefont {W.}~\bibnamefont {Burkett}}, \ and\
  \bibinfo {author} {\bibfnamefont {M.}~\bibnamefont {Xiao}},\ }\href@noop {}
  {\bibfield  {journal} {\bibinfo  {journal} {Opt. Lett.}\ }\textbf {\bibinfo
  {volume} {25}},\ \bibinfo {pages} {1732} (\bibinfo {year}
  {2000})}\BibitemShut {NoStop}%
\bibitem [{\citenamefont {Klempt}\ \emph {et~al.}(2006)\citenamefont {Klempt},
  \citenamefont {van Zoest}, \citenamefont {Henninger}, \citenamefont {Topic},
  \citenamefont {Rasel}, \citenamefont {Ertmer},\ and\ \citenamefont
  {Arlt}}]{Arlt06}%
  \BibitemOpen
  \bibfield  {author} {\bibinfo {author} {\bibfnamefont {C.}~\bibnamefont
  {Klempt}}, \bibinfo {author} {\bibfnamefont {T.}~\bibnamefont {van Zoest}},
  \bibinfo {author} {\bibfnamefont {T.}~\bibnamefont {Henninger}}, \bibinfo
  {author} {\bibfnamefont {O.}~\bibnamefont {Topic}}, \bibinfo {author}
  {\bibfnamefont {E.}~\bibnamefont {Rasel}}, \bibinfo {author} {\bibfnamefont
  {W.}~\bibnamefont {Ertmer}}, \ and\ \bibinfo {author} {\bibfnamefont
  {J.}~\bibnamefont {Arlt}},\ }\href {\doibase 10.1103/PhysRevA.73.013410}
  {\bibfield  {journal} {\bibinfo  {journal} {Phys. Rev. A}\ }\textbf {\bibinfo
  {volume} {73}},\ \bibinfo {pages} {013410} (\bibinfo {year}
  {2006})}\BibitemShut {NoStop}%
\bibitem [{\citenamefont {Kumar}\ \emph {et~al.}(2017)\citenamefont {Kumar},
  \citenamefont {Fan}, \citenamefont {K{\"u}bler}, \citenamefont {Sheng},\ and\
  \citenamefont {Shaffer}}]{kumar2017}%
  \BibitemOpen
  \bibfield  {author} {\bibinfo {author} {\bibfnamefont {S.}~\bibnamefont
  {Kumar}}, \bibinfo {author} {\bibfnamefont {H.}~\bibnamefont {Fan}}, \bibinfo
  {author} {\bibfnamefont {H.}~\bibnamefont {K{\"u}bler}}, \bibinfo {author}
  {\bibfnamefont {J.}~\bibnamefont {Sheng}}, \ and\ \bibinfo {author}
  {\bibfnamefont {J.~P.}\ \bibnamefont {Shaffer}},\ }\href@noop {} {\bibfield
  {journal} {\bibinfo  {journal} {Scientific Reports}\ }\textbf {\bibinfo
  {volume} {7}},\ \bibinfo {pages} {42981} (\bibinfo {year}
  {2017})}\BibitemShut {NoStop}%
\end{thebibliography}


\end{document}